\documentclass[a4paper, 12pt]{article}
\usepackage[textwidth=17cm]{geometry}
\usepackage{amscd}
\usepackage{amssymb}
\usepackage{amsmath}
\usepackage{graphicx,multirow}
\usepackage{array}
\usepackage{subfigure }
\usepackage[misc]{ifsym}
\usepackage{amsfonts}
\usepackage{latexsym}
\usepackage{color}
\usepackage{booktabs}
\usepackage{algorithm}
\usepackage{algorithmicx}
\usepackage{hyperref}

\usepackage{placeins}

\numberwithin{equation}{section}
\newtheorem{expl}{Example}[section]

\newtheorem{theorem}{Theorem}[section]
\newtheorem{lemma}[theorem]{Lemma}

\newtheorem{rem}[theorem]{Remark}

 \numberwithin{figure}{section}

\def\cL {{\cal L}}

\DeclareMathOperator{\Tr}{Tr}

\title{\Large\bf Full- and low-rank exponential midpoint schemes \\
for forward and adjoint Lindblad equations}
\author{\normalsize
Hao Chen\thanks{College of Mathematics Science, Chongqing Normal University, Chongqing, China.
Email address: hch@cqnu.edu.cn}
\and
\normalsize Alfio Borz\`{i}\thanks{Institut f\"{u}r Mathematik,
Universit\"{a}t W\"{u}rzburg, W\"{u}rzburg, Germany.
Email address: alfio.borzi@mathematik.uni-wuerzburg.de }
}

\date{}

\begin{document}

\maketitle

\vspace{-10pt}

\begin{abstract}
The Lindblad equation is a widely used quantum master equation to model the dynamical evolution of open quantum systems whose states are described by density matrices. This equation is also a fundamental building block to design optimal control functions. In this paper we develop full- and low-rank exponential midpoint integrators for solving both the forward and adjoint Lindblad equations. These schemes are applicable to optimize-then-discretize approaches for optimal control of open quantum systems. We show that the proposed schemes preserve positivity and trace unconditionally. Furthermore, convergence of these numerical schemes is proved theoretically and verified numerically.
\\

\end{abstract}


\vspace{-20pt}

\setlength\abovedisplayskip{4pt}
\setlength\belowdisplayskip{4pt}


\section{Introduction}
The optimal control of quantum systems has important applications in various fields, such as
NMR spectroscopy \cite{Fouquieres2011,Herbruggen2005,Khaneja2001,TOSNER2009},
quantum chemistry \cite{Khaneja2005,Maday2003,Reich2012,Zhu1998}, quantum information processing \cite{Doria2011,Egger2014} and molecular physics \cite{Palao2002}. We refer the reader to \cite{Alessandro2008,Wiseman2009} for a few references on mathematical tools developed for quantum optimal control. Optimal control problems for closed quantum systems have received significant attention in the past several decades and many numerical algorithms have been developed in the literature. Besides a monograph \cite{Borzi2017} on computational methods for
closed quantum control problems, we mention, among others, Gradient Ascent Pulse Engineering (GRAPE) \cite{Khaneja2005}, Chopped RAndom Basis (CRAB) algorithm \cite{Doria2011,Caneva2011,Rach2015}, Krotov method \cite{Reich2012,Krotov1995} and other monotonically converging gradient-based algorithms \cite{Gollub2008,Maday2003,Ohtsuki2004}.

Open quantum optimal control problems, where dissipation and dephasing effects enter the models, have also been studied; see, e.g., \cite{Goerz2014,Machnes2011,Schulte-Herbruggen2011,Wenin2008}.
The main difference between open and closed quantum systems is that the Schr\"{o}dinger equation  is replaced by a Markovian Lindblad master equation and the state vector by a density matrix \cite{Breuer}. For solving open quantum control problems, the most widely used algorithms are the open system versions of the GRAPE \cite{Schulte-Herbruggen2011}, CRAB \cite{Caneva2011,Rach2015} and Krotov algorithm \cite{Krotov1995}. For an open system of dimension $m$, a standard approach for optimal control is to reformulate the density matrix as a vector  of dimension $m^2\times 1$ and the Lindblad generator as a matrix of size $m^2\times m^2$. In this representation, time stepping
of the density matrix is usually obtained by matrix exponential of superoperators of dimension $m^2\times m^2$, which would be expensive even for moderate $m$. To address this issue, a natural idea is to propagate the density matrix directly through some integration methods; see, e.g., the combination of GRAPE and Runge-Kutta methods \cite{Boutin2017}. In addition, a gradient-based strategy combined with quantum trajectories and automatic differentiation has been studied in \cite{Abdelhafez2019}.

The motivation of this paper is threefold. First, it is well-known that the Lindblad master equation possesses semi-positiveness and trace preserving properties \cite{Gorini,Lindblad}. These properties of the density matrix are of fundamental physical significance, and whether they can be preserved at the discrete level is a crucial issue in numerical simulations. We note that there is very limited research being done on positivity preserving scheme for the Lindblad equation; and only a few works focusing on problems with time-independent Hamiltonian have been discussed in \cite{Appelo,Bidegaray,Cao,Riesch,Riesch1,Saut,Schlimgen,Ziolkowski} for the preservation of  positivity. The literature on positivity preserving scheme is more scarce for the Lindblad equation with time-dependent Hamiltonian, which appears typically in open quantum optimal control problems.

Second, most of the existing numerical methods for open quantum optimal control problems, such as GRAPE, require to solve and store state vectors of size $m^2\times 1$ or density matrices of size $m\times m$ at all the time grids. As the Hilbert space dimension $m$ increases, the optimizer would require expensive computational cost and memory.
To reduce the memory requirement and then the overall computational cost,
a potential strategy is to employ low-rank representation of the density matrix. In this setting, only matrices of size $m\times r$ with $r\ll m$ need to be computed and stored. Although there exist some low-rank schemes for solving the Lindblad equation \cite{Appelo,Chen3,LeBris1,LeBris2}, the use of low-rank algorithms for open quantum optimal control problems is still missing.

Third, we note that rigorous numerical analysis on low-rank schemes for open quantum systems is largely unavailable. The only numerical analysis of low-rank scheme for Lindblad equations we are aware of is found in \cite{Chen3}, which focuses on first-order exponential Euler scheme. To the best of our knowledge, no numerical analysis on low-rank algorithms for adjoint Lindblad equations is currently available.

All these facts motivate us to develop and analyze positivity and trace preserving, effective, and efficient numerical methods for solving forward and adjoint Lindblad equations, which appear typically in optimizer for optimal control problems of open quantum systems. For this purpose, we propose second-order full-rank exponential midpoint schemes for both the forward and adjoint Lindblad equations, followed by the low-rank variants of the exponential schemes. Positivity and trace preserving properties have been discussed and rigorous error estimates have been given for the proposed full- and low-rank algorithms.

This paper is organized as follows. In Section \ref{sec-preliminary}, we begin with some preliminaries, introducing the Lindblad master equation as well as the related
optimal control problems. In Section \ref{sec-ExpInt}, we introduce our full-rank and low-rank exponential midpoint schemes for discretizing the forward and adjoint Lindblad equations.
Sections \ref{sec-err1} and \ref{sec-err2} are devoted to the error analysis of the proposed exponential integrators for the forward and adjoint Lindblad equations, respectively.
In Section \ref{sec-experiments}, we report results of numerical experiments that
successfully validate our theoretical results.
A section of conclusion completes this work.

\section{Preliminary}\label{sec-preliminary}
The Lindblad equation is a widely used Markovian quantum master equation to model the dynamical evolution of open quantum systems \cite{Breuer,Davies1976}. In the case of quantum systems consisting of $K$ dephasing $d$-level qudits undergoing open quantum dynamics, the Lindblad equation is given by \cite{Gorini,Lindblad}:
\begin{equation}\label{LindbladEq}
  \dot{\rho}(t)=-i \, [H,\rho(t)]+\sum_{k=1}^K\gamma_k\left(L_k \, \rho(t) \, L_k^{\dag}-\frac{1}{2}\left\{L_k^{\dag}L_k,\rho(t)\right\}\right),
\end{equation}
where $\rho(t)\in \mathbb{C}^{m\times m}$ is the density matrix, describing the state of the system, being initially in the state $\rho(0)=\rho_0$, and $H=H^{\dag}$ is the Hamiltonian operator describing the unitary evolution of the qudit; $H$ can be time dependent. Further, the $L_k$ are the Lindblad or jump operators characterizing the dissipation channels, and $\gamma_k\geq0$ are the decay parameters for each of the $K$ channels. In \eqref{LindbladEq} and in the following, the superscripts $\dag$, $\top$ and $*$ denote the adjoint, transpose and complex conjugate operators, respectively.

On the other hand, we can define the operator $\cL$ as follows:
$$
\cL(\rho) := \sum_{k=1}^K\gamma_k\left(L_k \, \rho \, L_k^{\dag}-\frac{1}{2}\left\{L_k^{\dag}L_k,\rho \right\}\right),
$$
and write the Lindblad equation as: $\dot{\rho}(t)=-i \, [H,\rho(t)]+\cL(\rho(t))$.

For open quantum optimal control problem, one could consider two control mechanisms: Hamiltonian (coherent) control and environment (incoherent) control. In the first case, a typical coherent control is a shaped laser pulse that appears as a control Hamiltonian $H_c(t)=V \, u(t)$
(dipole approximation) as follows
$$
H(t)=H_0 + H_c(t),
$$
where $H_0$ represents the Hamiltonian of the uncontrolled system, $V$ denotes
a dipole interaction (moment) matrix, and $u$ denotes the control function.
In the second case, the action of the control is performed by changing the state
of the environment, which could be modelled as time-varying coefficients
$\gamma_k = \gamma_k(t)$ of the master equation; see, e.g., \cite{Pechen2006}.

We introduce the adjoint Hermitian function $q(t)\in \mathbb{C}^{m\times m}$ that
satisfies the adjoint Lindblad equation:
\begin{equation}\label{AdjointLindbladEq}
  \dot{q}(t)=-i \, [H,q(t)] - \sum_{k=1}^K\gamma_k\left(L_k^{\dag} \, q(t) \, L_k-\frac{1}{2}\left\{L_k^{\dag}L_k, q(t)\right\}\right) .
\end{equation}
In compact form, we can write: $\dot{q}(t)=-i \, [H,q(t)]  -  \cL^\dag (q(t))$.
Notice that the evolution modelled by \eqref{AdjointLindbladEq}
is backwards in time, with a terminal condition that results specified
by the choice of the cost functional.

In order to formulate the purpose and cost of the control, and write the Lagrange
function that allows a simple derivation of the adjoint Lindblad equation, we
recall the Hilbert-Schmidt scalar product $ \langle X,Y \rangle := \Tr (X^\dag  \, Y)$,
where $X$ and $Y$  are two Hilbert-Schmidt operators.

Now, suppose we wish to maximize the projection/overlap with a given
state $Q=Q^\dag$, then we would consider the functional
$$
J (\rho) :=  \Tr (Q \, \rho(T)) = \langle Q , \rho(T)  \rangle .
$$
In this case, the terminal condition for the adjoint variable is given by $q(T)=Q$.
In addition to maximizing the overlap, in the case of Hamiltonian control,
one could ask to minimize the $L^2(0,T)$
cost of the control, in which case the cost functional becomes
$$
J (\rho) :=  \langle Q , \rho(T)  \rangle  - \frac{\alpha}{2} \, \int_0^T u^2(t) \, dt
$$

We remark that the forward Lindblad equation \eqref{LindbladEq} preserves two important properties \cite{Gorini,Lindblad}:
if $\rho_0$ is a Hermitian and positive semidefinite matrix with unit trace, then $\rho(t)$ is Hermitian and positive semidefinite and has unit trace for all $t \ge 0$. Similar result also holds for the adjoint Lindblad equation \eqref{AdjointLindbladEq}. In fact, if we define $\tilde{q}(t):=q(T-t)$, then $\tilde{q}$ satisfies a forward Lindblad equation with initial condition $\tilde{q}(0)=Q$. Then $\tilde{q}(t)$ is Hermitian and positive semidefinite and has unit trace for all $t\in[0,T]$ if $Q$ is Hermitian and positive semidefinite with unit trace. We have the following result.

\begin{lemma}\label{PositivityTracePre}
 Assume that $\rho_0$ and $Q$ are Hermitian and positive semidefinite matrices with unit trace, then the solutions of the forward Lindblad equation \eqref{LindbladEq} and the backward Lindblad equation \eqref{AdjointLindbladEq} are both Hermitian and positive semidefinite and have unit trace for all $t\in[0,T]$.
\end{lemma}

Throughout this paper, we will always assume that $\rho_0$ and $Q$ are Hermitian and positive semidefinite matrices with unit trace.

We end this section with some notes on notations employed.
 If a matrix $\varrho\in \mathbb{C}^{m\times m}$ is Hermitian and positive semidefinite, we denote $\varrho\geq0$.
The trace norm of the matrix $\varrho$ is defined as $\|\varrho\|_1=\Tr(\sqrt{\varrho^{\dag}\varrho})=\sum_{j=1}^m\sigma_j(\varrho)$, where $\sigma_1(\varrho)\geq\sigma_2(\varrho)\geq\ldots\geq\sigma_m(\varrho)$ denote the singular values of $\varrho$. If $\varrho\geq0$, we have that $\|\varrho\|_1=\Tr(\varrho)$. We denote $\|\varrho\|_F$ the Frobenius norm of $\varrho$.

\section{Exponential integrators}\label{sec-ExpInt}
Note that one needs to solve the forward and backward Lindblad equations and store their solutions repeatedly in gradient-based algorithms for optimal control of open quantum systems. So it is desirable to design numerical schemes with low computational cost and storage for these Lindblad equations.
In this section, we develop full- and low-rank exponential integrators to solve the Lindblad equation \eqref{LindbladEq} and its adjoint \eqref{AdjointLindbladEq}.
It is convenient to introduce the operator
\begin{equation}\label{equ2.1}
  A(t):=- i \, H(t)-\frac{1}{2}\sum_{k=1}^K\gamma_k(t) \, L_k^{\dag} \, L_k.
\end{equation}
Then, we can rewrite \eqref{LindbladEq} as
\begin{equation}\label{equ2.2}
  \dot{\rho}(t)=A(t) \, \rho(t)+\rho(t) \,A^{\dag}(t)+\sum_{k=1}^K\gamma_k(t) \, L_k \, \rho(t) \,L_k^{\dag}, \qquad \rho(0)=\rho_0,
\end{equation}
and the adjoint Lindblad equation \eqref{AdjointLindbladEq} can be written as
\begin{equation}\label{equ2.3}
  \dot{q}(t)=-A^{\dag}(t) \, q(t) - q(t) \,A(t)-\sum_{k=1}^K\gamma_k(t) \, L_k^{\dag} \, q(t) \,L_k, \qquad q(T)=Q.
\end{equation}
Note that the forward equation \eqref{equ2.2} and the backward equation \eqref{equ2.3} cover both cases of coherent control ($\gamma_k(t)\equiv\gamma_k$) and incoherent control ($H(t)\equiv H_0$).

\subsection{Full-rank exponential integrators}
Let us first consider a full-rank exponential midpoint scheme for the forward problem \eqref{equ2.2}. We discretize the time interval $[0,T]$ by the uniform grid $\{t_n\}_{n=0}^N$
with time step $\tau=T/N$, and seek a numerical approximation $\rho_n$ of the exact solution $\rho(t_n)$. Let $A_j=A(t_j)$ and  the forward problem
\eqref{equ2.2} can be rewritten as
\begin{equation}\label{equ2.4}
  \dot{\rho}(t)=A_j \, \rho(t)+\rho(t) \,A^{\dag}_j+F(t,\rho(t),A_j), \qquad \rho(0)=\rho_0,
\end{equation}
where
\[F(t,\rho(t),A_j)=\sum_{k=1}^K\gamma_k(t) \, L_k \, \rho(t) \,L_k^{\dag}+(A(t)-A_j)\, \rho(t)+\rho(t)(A^{\dag}(t) -A^{\dag}_j).\]
Integrating \eqref{equ2.4} from $t_n$ to $t$ and applying the variation-of-constants formula, we get
\begin{equation}\label{equ2.5}
  \rho(t)=e^{(t-t_n)A_j} \, \rho(t_n) \, e^{(t-t_n)A^{\dag}_j} +
  \int_0^{t-t_n}e^{(t-t_n-s)A_j} \, F(t_n+s, \rho(t_n+s), A_j) \, e^{(t-t_n-s)A^{\dag}_j}ds.
\end{equation}
Letting $t=t_n+\frac{\tau}{2}:=t_{n+1/2}$ (resp. $t=t_{n+1}$) and $j=n$ (resp. $j=n+1/2$) in \eqref{equ2.5}, we obtain
\begin{subequations}\label{equ2.6}
\begin{align}
  &\rho(t_{n+1/2})=e^{\frac{\tau}{2} A_n} \, \rho(t_n) \, e^{\frac{\tau}{2} A_n^{\dag}} +
  \int_0^{\frac{\tau}{2}}e^{(\frac{\tau}{2}-s)A_n} \, F(t_n+s, \rho(t_n+s), A_n) \, e^{(\frac{\tau}{2}-s)A_n^{\dag}}ds,\\
  &\rho(t_{n+1})=e^{\tau A_{n+1/2}} \, \rho(t_n) \, e^{\tau A_{n+1/2}^{\dag}} +
  \int_0^{\tau}e^{(\tau-s)A_{n+1/2}} \, F(t_n+s, \rho(t_n+s), A_{n+1/2}) \, e^{(\tau-s)A_{n+1/2}^{\dag}}ds.
\end{align}
\end{subequations}
Approximating the integrals in \eqref{equ2.6} by left-rectangle quadrature formula  and
midpoint quadrature formula, respectively, we get the full-rank exponential midpoint (FREM) scheme
\begin{subequations}\label{equ2.7}
\begin{align}
  &\rho_{n+1/2}=e^{\frac{\tau}{2} A_n} \left( \rho_n + \frac{\tau}{2}\sum_{k=1}^K\gamma_k(t_n) \, L_k \, \rho_n \,L_k^{\dag} \right) e^{\frac{\tau}{2} A_n^{\dag}} ,\label{equ2.7a}\\
  &\rho_{n+1}=e^{\tau A_{n+1/2}} \, \rho_n \, e^{\tau A_{n+1/2}^{\dag}} +
  \tau  \sum_{k=1}^K\gamma_k(t_{n+1/2})e^{\frac{\tau}{2} A_{n+1/2}} \, L_k \, \rho_{n+1/2} \,L_k^{\dag} \, e^{\frac{\tau}{2} A_{n+1/2}^{\dag}},\label{equ2.7b}
\end{align}
\end{subequations}
$n=0,\ldots,N-1$.

Now we develop a full-rank exponential midpoint scheme for the backward problem \eqref{equ2.3}.
Similarly, we can rewrite \eqref{equ2.3} as
\begin{equation}\label{equ2.8}
  \dot{q}(t)=-A^{\dag}_j \, q(t)-q(t) \,A_j-\tilde{F}(t,q(t),A_j), \qquad q(T)=Q,
\end{equation}
where
\[\tilde{F}(t,q(t),A_j)=\sum_{k=1}^K\gamma_k(t) \, L_k^{\dag} \, q(t) \,L_k+(A^{\dag}(t)-A^{\dag}_j)\, q(t)+q(t)(A(t) -A_j).\]
Integrating \eqref{equ2.8} from $t_{n+1}$ to $t$ and applying the variation-of-constants formula, we obtain
\begin{equation}\label{equ2.9}
  q(t)=e^{(t_{n+1}-t)A^{\dag}_j} \, q(t_{n+1}) \, e^{(t_{n+1}-t)A_j} +
  \int_{t-t_n}^{\tau}e^{(t_n+s-t)A^{\dag}_j} \, \tilde{F}(t_n+s, q(t_n+s), A_j) \, e^{(t_n+s-t)A_j}ds.
\end{equation}
Taking $t=t_{n+1/2}$ (resp. $t=t_{n}$) and $j=n+1$ (resp. $j=n+1/2$) in \eqref{equ2.9}, we get
\begin{subequations}\label{equ2.10}
\begin{align}
  &q(t_{n+1/2})=e^{\frac{\tau}{2} A_{n+1}^{\dag}} \, q(t_{n+1}) \, e^{\frac{\tau}{2} A_{n+1}} +
  \int_{\frac{\tau}{2}}^{\tau}e^{(s-\frac{\tau}{2})A_{n+1}^{\dag}} \, \tilde{F}(t_n+s, q(t_n+s), A_{n+1}) \, e^{(s-\frac{\tau}{2})A_{n+1}}ds,\\
  &q(t_{n})=e^{\tau A_{n+1/2}^{\dag}} \, q(t_{n+1}) \, e^{\tau A_{n+1/2}} +
  \int_0^{\tau}e^{sA_{n+1/2}^{\dag}} \, \tilde{F}(t_n+s, q(t_n+s), A_{n+1/2}) \, e^{sA_{n+1/2}}ds.
\end{align}
\end{subequations}
Approximating the integrals in \eqref{equ2.10} by right-rectangle quadrature formula  and
midpoint quadrature formula, respectively, we obtain the FREM scheme for the backward Lindblad equation
\begin{subequations}\label{equ2.11}
\begin{align}
  &q_{n+1/2}=e^{\frac{\tau}{2} A_{n+1}^{\dag}} \left( q_{n+1} + \frac{\tau}{2}\sum_{k=1}^K\gamma_k(t_{n+1}) \, L_k^{\dag} \, q_{n+1} \,L_k \right) e^{\frac{\tau}{2} A_{n+1}} ,\label{equ2.11a}\\
  &q_{n}=e^{\tau A_{n+1/2}^{\dag}} \, q_{n+1} \, e^{\tau A_{n+1/2}} +
  \tau \sum_{k=1}^K\gamma_k(t_{n+1/2}) \,e^{\frac{\tau}{2} A_{n+1/2}^{\dag}} \, L_k^{\dag} \, q_{n+1/2} \,L_k \, e^{\frac{\tau}{2} A_{n+1/2}},\label{equ2.11b}
\end{align}
\end{subequations}
$n=N-1,\ldots,0$. Note that $q_n$ and $q_{n+1/2}$ are approximations to $q(t_n)$ and $q(t_{n+1/2})$, respectively.

In order to be concise, we simply denote the FREM scheme \eqref{equ2.7} (resp. \eqref{equ2.11}) as the map $\rho_{n+1}=\Phi(t_n,\rho_n)$ (resp. $q_n=\Psi(t_{n+1},q_{n+1})$). Note that the main computational cost of the FREM scheme \eqref{equ2.7} (or \eqref{equ2.11}) is due to the $6m$
matrix-vector products associated matrix exponential at every time step.

\begin{rem}
 Assume that the initial value $\rho_0$ of the forward problem \eqref{equ2.2} and the terminal value $Q$ of the backward problem \eqref{equ2.3} are both Hermitian and positive semidefinite. Then, it is easy to show that for any time step size $\tau>0$, the FREM schemes \eqref{equ2.7} and \eqref{equ2.11} preserve the Hermitian and positive semidefinite property, i.e. $\rho_n$ and $q_n$ are Hermitian and positive semidefinite for all $n=0,\ldots,N$. This can be proved by induction and noting that each term on the right-hand side of the scheme \eqref{equ2.7} (and \eqref{equ2.11}) is Hermitian and positive semidefinite.
\end{rem}

We also remark that the FREM schemes \eqref{equ2.7} and \eqref{equ2.11} might not preserve the unit trace of the density matrices. In order to preserve unit trace of the density matrices, we
propose the normalized FREM schemes
\begin{subequations}\label{equ2.12}
  \begin{align}
    & \tilde{\rho}_{n+1}=\Phi(t_n,\rho_n),\qquad \rho_{n+1}=\frac{\tilde{\rho}_{n+1}}{\Tr(\tilde{\rho}_{n+1})},\qquad n=0,\ldots,N-1,\label{equ2.12a}\\
    &\tilde{q}_n=\Psi(t_{n+1},q_{n+1}),\qquad q_n=\frac{\tilde{q}_n}{\Tr(\tilde{q}_n)},\qquad \qquad n=N-1,\ldots,0.\label{equ2.12b}
  \end{align}
\end{subequations}

\subsection{Low-rank exponential integrators}
Now we consider the low-rank variants of the FREM schemes \eqref{equ2.7} and \eqref{equ2.11}. Our aim is to reduce the computational cost while at the same time retain the accuracy of the underlying FREM scheme. The idea is to seek and do computations on factors $X_n\in \mathbb{C}^{m\times r_n}$ (resp. $Y_n\in \mathbb{C}^{m\times \tilde{r}_n}$) with $r_n\ll m$ (resp. $\tilde{r}_n\ll m$) instead of $\rho_n$ (resp. $q_n$) such that the solutions of the forward and backward Lindblad equations can be well approximated as
\[\rho(t_n)\approx X_nX_n^{\dag}:=\varrho_n,\]
and
\[q(t_n)\approx Y_nY_n^{\dag}:=p_n,\]
respectively, where we denote with $\varrho_n$ (resp. $p_n$) the numerical low-rank solution to the forward (resp. backward) Lindblad equation in order to distinguish it from $\rho_n$ (resp. $q_n$), the full-rank numerical solution of the same equation.

Now assume that $\rho_n=X_nX_n^{\dag}$ and $\rho_{n+1/2}=X_{n+1/2}X_{n+1/2}^{\dag}$ and inserting these factorizations into \eqref{equ2.7} yields
\begin{subequations}\label{equ2.13}
  \begin{align}
  &G_{n}=\left[\sqrt{\tau\gamma_1(t_n)}L_1X_{n},\ldots,\sqrt{\tau\gamma_K(t_n)}L_KX_{n}\right],\label{equ2.13a}\\
  &X_{n+1/2}=e^{\frac{\tau}{2} A_n}\left[X_n, \sqrt{0.5}G_{n}\right],\label{equ2.13b}\\
  &G_{n+1/2}=\left[\sqrt{\tau\gamma_1(t_{n+1/2})}L_1X_{n+1/2},\ldots,\sqrt{\tau\gamma_K(t_{n+1/2})}L_KX_{n+1/2}\right],\label{equ2.13c}\\
  &X_{n+1}=\left[e^{\tau A_{n+1/2}}X_n, \, e^{\frac{\tau}{2} A_{n+1/2}}G_{n+1/2}\right].\label{equ2.13d}
\end{align}
\end{subequations}
By the notation in \eqref{equ2.13a} we mean that the $K$ matrices $\sqrt{\tau\gamma_k(t_n)}L_kX_{n}$ are placed side by side.

We remark that for many problems, the exact matrix exponential $e^{\frac{\tau}{2} A_n}$ or the exact value of the product of matrix exponential times vectors $e^{\frac{\tau}{2} A_n}X_n$  may be costly to compute and approximations may be
required. In our low-rank algorithms we will denote by $\mathfrak{e}^{\frac{\tau}{2} A_n}$ (resp. $\mathfrak{e}^{\frac{\tau}{2} A_n}X_n$) an approximation of $e^{\frac{\tau}{2} A_n}$ (resp. $e^{\frac{\tau}{2} A_n}X_n$).

In addition, note that matrices $G_{n}$ and $X_{n+1}$ have much more columns than $X_n$. Better approximations can be obtained by applying column compression techniques to these factors. This can be computed by truncating the singular value decomposition (SVD) of the given matrix. We denote with $\mathcal{T}_{\varepsilon_1}(\cdot)$ the truncated SVD of a matrix with error tolerance $\varepsilon_1>0$ in the sense that $\mathcal{T}_{\varepsilon_1}(X)$ represents the best rank $r$ approximation of the matrix $X\in \mathbb{C}^{m\times s}$ in Frobenius norm, where $r$ is the minimal integer such that $\sum_{j=r+1}^s\sigma^2_{j}(X)\leq \varepsilon_1$. We then get
\begin{equation}\label{equ2.14}
  \left\|XX^{\dag}-\mathcal{T}_{\varepsilon_1}(X)\mathcal{T}_{\varepsilon_1}(X)^{\dag}\right\|_1=
  \sum_{j=r+1}^s\sigma^2_{j}(X)\leq \varepsilon_1.
\end{equation}

Now, given initial low-rank approximation $\rho_0\approx \varrho_0=X_0X_0^{\dag}$ with $X_0\in \mathbb{C}^{m\times r_0}$ and $\Tr(\varrho_0)=1$, we define one step of the low-rank exponential midpoint (LREM) scheme as
\begin{subequations}\label{equ2.15}
  \begin{align}
  &\tilde{G}_{n}=\left[\sqrt{\tau\gamma_1(t_n)}L_1X_{n},\ldots,\sqrt{\tau\gamma_K(t_n)}L_KX_{n}\right], \qquad G_n = \mathcal{T}_{\varepsilon_1}(\tilde{G}_n),\\
  &\tilde{X}_{n+1/2}=\mathfrak{e}^{\frac{\tau}{2} A_n}\left[X_n, \sqrt{0.5}G_{n}\right],\qquad X_{n+1/2}=\mathcal{T}_{\varepsilon_1}(\tilde{X}_{n+1/2}),\\
  &\tilde{G}_{n+1/2}=\left[\sqrt{\tau\gamma_1(t_{n+1/2})}L_1X_{n+1/2},\ldots,\sqrt{\tau\gamma_K(t_{n+1/2})}L_KX_{n+1/2}\right],\\
  &G_{n+1/2}=\mathcal{T}_{\varepsilon_1}(\tilde{G}_{n+1/2}),\qquad \tilde{X}_{n+1}=\left[\mathfrak{e}^{\tau A_{n+1/2}}X_n, \, \mathfrak{e}^{\frac{\tau}{2} A_{n+1/2}}G_{n+1/2}\right],\\
  &\hat{X}_{n+1}=\mathcal{T}_{\varepsilon_1}(\tilde{X}_{n+1}),
  \qquad X_{n+1}=\frac{\hat{X}_{n+1}}{\|\hat{X}_{n+1}\|_F},
\end{align}
\end{subequations}
$n=0,\ldots,N-1$.

\begin{rem}
Note that $\varrho_{n+1}=X_{n+1}X_{n+1}^{\dag}$ and it follows that the LREM scheme \eqref{equ2.15} is positivity and trace preserving, i.e.
\end{rem}
\begin{equation}\label{equ2.16}
  \|\varrho_{n+1}\|_1=\Tr(X_{n+1}X_{n+1}^{\dag})=
  \frac{\Tr(\hat{X}_{n+1}\hat{X}_{n+1}^{\dag})}{\|\hat{X}_{n+1}\|_F^2}=1,\qquad n=0,\ldots,N-1.
\end{equation}

We also remark that the LREM scheme \eqref{equ2.15} is equivalent to
\begin{subequations}\label{equ2.17}
  \begin{align}
  &\tilde{\varrho}_{n+1}=\Phi(t_n,\varrho_n),\label{equ2.17a}\\
  &\hat{\varrho}_{n+1}=\tilde{\varrho}_{n+1}-\vartheta_{n+1},\label{equ2.17b}\\
  & \varrho_{n+1}=\frac{\hat{\varrho}_{n+1}}{\Tr(\hat{\varrho}_{n+1})},\label{equ2.17c}
\end{align}
\end{subequations}
$n=0,\ldots,N-1$, where $\hat{\varrho}_{n+1}=\hat{X}_{n+1}\hat{X}_{n+1}^{\dag}$ and the matrix $\vartheta_{n+1}$ can be seen as the perturbation caused by the approximations to the matrix exponential times vectors and the column compression procedures.

Similarly, given terminal low-rank approximation $Q\approx p_N=Y_NY_N^{\dag}$ with $Y_N\in \mathbb{C}^{m\times \tilde{r}_N}$ and $\Tr(p_N)=1$, we can get the LREM scheme for the backward Lindblad equation as
\begin{subequations}\label{equ2.18}
  \begin{align}
  &\tilde{W}_{n+1}=\left[\sqrt{\tau\gamma_1(t_{n+1})}L_1^{\dag}Y_{n+1},\ldots,\sqrt{\tau\gamma_K(t_{n+1})}L_K^{\dag}Y_{n+1}\right], ~ W_{n+1} = \mathcal{T}_{\varepsilon_1}(\tilde{W}_{n+1}),\\
  &\tilde{Y}_{n+1/2}=\mathfrak{e}^{\frac{\tau}{2} A_{n+1}^{\dag}}\left[Y_{n+1}, \sqrt{0.5}W_{n+1}\right],\qquad Y_{n+1/2}=\mathcal{T}_{\varepsilon_1}(\tilde{Y}_{n+1/2}),\\
  &\tilde{W}_{n+1/2}=\left[\sqrt{\tau\gamma_1(t_{n+1/2})}L_1^{\dag}Y_{n+1/2},\ldots,\sqrt{\tau\gamma_K(t_{n+1/2})}L_K^{\dag}Y_{n+1/2}\right],\\
  &W_{n+1/2}=\mathcal{T}_{\varepsilon_1}(\tilde{W}_{n+1/2}),\qquad \tilde{Y}_{n}=\left[\mathfrak{e}^{\tau A_{n+1/2}^{\dag}}Y_{n+1}, \, \mathfrak{e}^{\frac{\tau}{2} A_{n+1/2}^{\dag}}W_{n+1/2}\right],\\
  &\hat{Y}_{n}=\mathcal{T}_{\varepsilon_1}(\tilde{Y}_{n}),
  \qquad Y_{n}=\frac{\hat{Y}_{n}}{\|\hat{Y}_{n}\|_F},
\end{align}
\end{subequations}
$n=N-1,\ldots,0$.

\begin{rem}
Note that $p_{n}=Y_{n}Y_{n}^{\dag}$ and it follows that the LREM scheme \eqref{equ2.18} is positivity and trace preserving, i.e.
\end{rem}
\begin{equation}\label{equ2.19}
  \|p_{n}\|_1=\Tr(Y_{n}Y_{n}^{\dag})=
  \frac{\Tr(\hat{Y}_{n}\hat{Y}_{n}^{\dag})}{\|\hat{Y}_{n}\|_F^2}=1,\qquad n=N-1,\ldots,0.
\end{equation}

We remark that the LREM scheme \eqref{equ2.19} is equivalent to
\begin{subequations}\label{equ2.20}
  \begin{align}
  &\tilde{p}_{n}=\Psi(t_{n+1},p_{n+1}),\label{equ2.20a}\\
  &\hat{p}_{n}=\tilde{p}_{n}-\theta_{n},\label{equ2.20b}\\
  & p_{n}=\frac{\hat{p}_{n}}{\Tr(\hat{p}_{n})},\label{equ2.20c}
\end{align}
\end{subequations}
$n=N-1,\ldots,0$, where $\hat{p}_{n}=\hat{Y}_{n}\hat{Y}_{n}^{\dag}$ and the matrix $\theta_{n}$ is the perturbation due to the approximations to the matrix exponential times vectors and the column compression procedures.

\section{Error analysis for forward problem}\label{sec-err1}
In this section, we perform error analysis of the proposed FREM scheme \eqref{equ2.12a} and LREM scheme \eqref{equ2.15} for the forward Lindblad equation. In the proofs, we will use the following result.
\begin{lemma}\label{lem3.1}
  (see \cite{Chen3}) For any Hermitian matrix $\sigma\in \mathbb{C}^{m\times m}$, it holds that
  \begin{equation*}
    \left\|e^{tA(s)}\sigma e^{tA(s)^{\dag}}\right\|_1\leq \left\|\sigma\right\|_1,\qquad \forall t\geq0,~s\in[0,T].
  \end{equation*}
\end{lemma}

\subsection{Error estimate of the FREM scheme}
First we perform consistency analysis of the FREM scheme \eqref{equ2.7}. Considering \eqref{equ2.6} and using the consistency of the left-rectangle quadrature formula and midpoint quadrature formula, we have
\begin{subequations}\label{equ3.1}
\begin{align}
  &\rho(t_{n+1/2})=e^{\frac{\tau}{2} A_n} \left( \rho(t_n) + \frac{\tau}{2}\sum_{k=1}^K\gamma_k(t_n) \, L_k \, \rho(t_n) \,L_k^{\dag} \right) e^{\frac{\tau}{2} A_n^{\dag}} +\mathcal{O}(\tau^2),\label{equ3.1a}\\
  &\rho(t_{n+1})=e^{\tau A_{n+1/2}} \, \rho(t_n) \, e^{\tau A_{n+1/2}^{\dag}} +
  \tau \sum_{k=1}^K\gamma_k(t_{n+1/2}) \,e^{\frac{\tau}{2} A_{n+1/2}} \, L_k \, \rho(t_{n+1/2}) \,L_k^{\dag} \, e^{\frac{\tau}{2} A_{n+1/2}^{\dag}}+\mathcal{O}(\tau^3).\label{equ3.1b}
\end{align}
\end{subequations}
Inserting \eqref{equ3.1a} into \eqref{equ3.1b} yields
\begin{equation}\label{equ3.2}
  \rho(t_{n+1})=\Phi(t_n,\rho(t_n)) + R_{n+1},
\end{equation}
and the truncation error $R_{n+1}$ satisfies
\begin{equation}\label{equ3.3}
  \max_{0\leq n\leq N-1}\|R_{n+1}\|_1\leq C_1 \, \tau^3,
\end{equation}
where the positive constant $C_1$ depends on $L_k$, $A(t)$, $\gamma_k(t)$, $\rho(t)$ and their first and second order derivatives.

\begin{lemma}\label{lem3.2}
 For any Hermitian matrices $\rho_n,\varrho_n\in \mathbb{C}^{m\times m}$, it holds that
 \[ \|\Phi(t_n,\rho_n)-\Phi(t_n,\varrho_n)\|_1\leq (1+\tau C_2+\tau^2C_2^2/2) \, \|\rho_n-\varrho_n\|_1,\qquad n=0,\ldots,N-1,\]
 where $C_2=\sum\limits_{k=1}^K\left(\max\limits_{0\leq t\leq T}\gamma_k(t)\right)\|L_k\|_1^2$.
\end{lemma}
\begin{proof}
We need to consider a single step of the FREM method \eqref{equ2.7}, applied at $t_n$ to the initial matrices $\rho_n$ and $\varrho_n$. We denote the intermediate values by $\rho_{n+1/2}$
and $\varrho_{n+1/2}$, respectively.
  Considering the difference of the equations \eqref{equ2.7a} with respect to different initial values $\rho_n$ and $\varrho_n$ and using Lemma \ref{lem3.1}, one obtains
  \begin{eqnarray*}
    \|\rho_{n+1/2}-\varrho_{n+1/2}\|_1&\leq & \|\rho_n-\varrho_n\|_1+\frac{\tau }{2} \sum_{k=1}^{K}\gamma_k(t_n)\|L_k(\rho_{n}-\varrho_{n})L_k^{\dag}\|_1\\
    &\leq& (1+\frac{\tau}{2}C_2)\|\rho_n-\varrho_n\|_1.
  \end{eqnarray*}
  Similarly, using \eqref{equ2.7b}, Lemma \ref{lem3.1}, and noting that $\rho_{n+1}=\Phi(t_n,\rho_n)$ and $\varrho_{n+1}=\Phi(t_n,\varrho_n)$, we have
  \begin{eqnarray*}
   \|\Phi(t_n,\rho_n)-\Phi(t_n,\varrho_n)\|_1 &=& \|\rho_{n+1}-\varrho_{n+1}\|_1\\
   &\leq& \|\rho_n-\varrho_n\|_1+\tau \sum_{k=1}^{K}\gamma_k(t_{n+1/2})\|L_k(\rho_{n+1/2}-\varrho_{n+1/2})L_k^{\dag}\|_1\\
    &\leq&\|\rho_n-\varrho_n\|_1+\tau C_2 \|\rho_{n+1/2}-\varrho_{n+1/2}\|_1\\
    &=&(1+\tau C_2+\tau^2 C_2^2/2)\|\rho_{n}-\varrho_{n}\|_1,
  \end{eqnarray*}
  which completes the proof.
\end{proof}

\begin{lemma}\label{lem3.3}
  Let $\sigma$ be Hermitian and positive semidefinite with unit trace, then it holds that
  \[1-C_1 \, \tau^3\leq\Tr(\Phi(t_n,\sigma))\leq 1+C_1 \, \tau^3,\qquad n=0,\ldots,N-1.\]
\end{lemma}
\begin{proof}
  Let $\varrho(t)$ be the solution of the Lindblad equation \eqref{equ2.2} with initial condition
  $\varrho(t_n)=\sigma$. It then follows that $\varrho(t)$ is Hermitian and positive semidefinite with unit trace, i.e., $\|\varrho(t)\|_1=\Tr(\varrho(t))=1$ for all $t\geq t_n$. Note that
  $\Phi(t_n,\sigma)$ is the numerical approximation to $\varrho(t_{n+1})$ by using the FREM scheme \eqref{equ2.7} for a single step with exact initial value $\varrho(t_n)$.
  By the consistency \eqref{equ3.2}-\eqref{equ3.3} of the FREM scheme \eqref{equ2.7}, we obtain that
  \[\|\Phi(t_n,\sigma)-\varrho(t_{n+1})\|_1\leq C_1 \, \tau^3.\]
  Using
 \[ \left|\|\Phi(t_n,\sigma)\|_1-\|\varrho(t_{n+1})\|_1\right|\leq\|\Phi(t_n,\sigma)-\varrho(t_{n+1})\|_1,\]
 we get
 \[\|\varrho(t_{n+1})\|_1-\|\Phi(t_n,\sigma)-\varrho(t_{n+1})\|_1\leq\|\Phi(t_n,\sigma)\|_1\leq
 \|\varrho(t_{n+1})\|_1+\|\Phi(t_n,\sigma)-\varrho(t_{n+1})\|_1.\]
  The desired result then follows from the positivity preserving property of the FREM scheme \eqref{equ2.7} and $\|\varrho(t_{n+1})\|_1=\Tr(\varrho(t_{n+1}))=1$.
\end{proof}

Now we present the error estimate for the numerical solution derived from the unnormalized FREM scheme \eqref{equ2.7} for the forward Lindblad equation \eqref{equ2.2}.

\begin{theorem}\label{thm3.1}
  The numerical solution $\rho_n$ generated by the unnormalized FREM scheme \eqref{equ2.7} with $\rho_0=\rho(0)$ satisfies the error estimate
  \[\|\rho(t_n)-\rho_n\|_1\leq C_3 \, \tau^2,\qquad 0\leq n\leq N,\]
  where the constant $C_3>0$ depends on $C_1,~C_2,~T$ but is independent of $\tau$ and $n$.
\end{theorem}
\begin{proof}
  Considering the difference between $\rho_{n+1}=\Phi(t_n,\rho_n)$ and \eqref{equ3.2}, and using \eqref{equ3.3} and Lemma \ref{lem3.2}, we obtain
  \begin{eqnarray*}
    \|\rho(t_{n+1})-\rho_{n+1}\|_1&\leq& \|\Phi(t_n,\rho(t_n))-\Phi(t_n,\rho_n)\|_1+\|R_{n+1}\|_1 \\
    &\leq & (1+\tau C_2+\tau^2C_2^2/2)\|\rho(t_n)-\rho_n\|_1+C_1\tau^3.
  \end{eqnarray*}
  By recursion, we obtain
  \begin{eqnarray*}
    \|\rho(t_{n})-\rho_{n}\|_1\leq (1+\tau C_2+\tau^2C_2^2/2)^n\|\rho(t_0)-\rho_0\|_1
    +C_1\tau^3\sum_{j=0}^{n-1}(1+\tau C_2+\tau^2C_2^2/2)^j.
  \end{eqnarray*}
  Noting that $\rho(t_0)-\rho_0=0$, we have
  \begin{eqnarray*}
    \|\rho(t_{n})-\rho_{n}\|_1\leq \frac{C_1}{C_2}(e^{C_2t_n}-1)\tau^2,
  \end{eqnarray*}
  and the desired result follows with $C_3:=\frac{C_1}{C_2}(e^{C_2T}-1)$.
\end{proof}

Now we are in the position to prove the convergence of the normalized FREM scheme \eqref{equ2.12a}.

\begin{theorem}\label{thm3.2}
    The numerical solution $\rho_n$ generated by the normalized FREM scheme \eqref{equ2.12a} with $\rho_0=\rho(0)$ satisfies the error estimate
  \[\|\rho(t_n)-\rho_n\|_1\leq 2C_3 \, \tau^2,\qquad 0\leq n\leq N,\]
  where the constant $C_3$ is as defined in Theorem \ref{thm3.1}.
  \end{theorem}
  \begin{proof}
    We denote $\hat{\rho}_{n+1}=\Phi(t_n,\hat{\rho}_n)$ with $\hat{\rho_0}=\rho(0)$, that means that $\hat{\rho}_n$ is the numerical solution generated by the unnormalized FREM scheme \eqref{equ2.7}.
    By Theorem \ref{thm3.1}, we have
   \begin{eqnarray*}
     \|\rho(t_{n+1})-\rho_{n+1}\|_1&\leq&\|\rho(t_{n+1})-\hat{\rho}_{n+1}\|_1+\|\hat{\rho}_{n+1}-\rho_{n+1}\|_1\\
     &\leq&C_3\tau^2+\|\hat{\rho}_{n+1}-\rho_{n+1}\|_1.
   \end{eqnarray*}
   Using Lemmas \ref{lem3.2} and \ref{lem3.3} and noting that $\rho_n\geq0$ and $\|\rho_n\|_1=1$, we obtain
   \begin{eqnarray*}
     \|\hat{\rho}_{n+1}-\rho_{n+1}\|_1&\leq&\|\hat{\rho}_{n+1}-\tilde{\rho}_{n+1}\|_1+\|\tilde{\rho}_{n+1}-\rho_{n+1}\|_1\\
     &=& \|\Phi(t_n,\hat{\rho}_n)-\Phi(t_n,\rho_n)\|_1+\|\rho_{n+1}(\Tr(\tilde{\rho}_{n+1})-1)\|_1\\
     &\leq& (1+\tau C_2+\tau^2C_2^2/2)\|\hat{\rho}_n-\rho_n\|_1+|\Tr(\Phi(t_n,\rho_n))-1|\cdot\|\rho_{n+1}\|_1\\
     &\leq&(1+\tau C_2+\tau^2C_2^2/2)\|\hat{\rho}_n-\rho_n\|_1+C_1\tau^3.
   \end{eqnarray*}
   By recursion and noting that $\hat{\rho}_0=\rho_0=\rho(0)$, we obtain
  \begin{eqnarray*}
    \|\hat{\rho}_{n+1}-\rho_{n+1}\|_1&\leq&(1+\tau C_2+\tau^2C_2^2/2)^{(n+1)}\|\hat{\rho}_0-\rho_0\|_1
    +C_1\tau^3\sum_{j=0}^{n}(1+\tau C_2+\tau^2C_2^2/2)^{j}\\
    &\leq& C_3\tau^2,
  \end{eqnarray*}
  which completes the proof.
  \end{proof}

\subsection{Error estimate of the LREM scheme}
Now we consider error estimate of the proposed LREM scheme \eqref{equ2.15} for the forward Lindblad equation. First, we analyze the bound of perturbation $\vartheta_{n+1}$ (defined in \eqref{equ2.17b}) in the following lemma.

\begin{lemma}\label{lem3.4}
 Let $\varepsilon_1>0$ be the error tolerance of the column compression algorithm used in the LREM scheme \eqref{equ2.15}. Assume that the matrix exponential algorithm used in the LREM scheme \eqref{equ2.15} satisfies $\|e^{\mu\tau A_{\nu}}\sigma e^{\mu\tau A_{\nu}^{\dag}}-\mathfrak{e}^{\mu\tau A_{\nu}}\sigma \mathfrak{e}^{\mu\tau A_{\nu}^{\dag}}\|_1\leq C_e\varepsilon_2\|\sigma\|_1$ for $\mu=0.5, 1$, $\nu=n,n+1/2$ with $n=0,\ldots,N-1$ and any $\sigma\in \mathbb{C}^{m\times m}$, where $0<\varepsilon_2<1$ is the
   corresponding error tolerance and $C_e>0$ is the error constant. Then it holds that
  \begin{equation*}
  \|\vartheta_{n+1}\|_1\leq \tilde{c}_1 \, \varepsilon_1 + \tilde{c}_2  \, \varepsilon_2,\qquad n=0,1,\ldots,N-1,
\end{equation*}
where the positive constants $\tilde{c}_1,~\tilde{c}_2$ depend on $C_e$, $C_2$ and $T$ but is independent of $\tau$, $n$, $\varepsilon_1$ and $\varepsilon_2$.
\end{lemma}
\begin{proof}
  Let us first define
  \begin{equation*}
    \varphi_{n+1}=\mathfrak{e}^{\tau A_{n+1/2}}\varrho_n\mathfrak{e}^{\tau A_{n+1/2}^{\dag}}+
    \mathfrak{e}^{\frac{\tau}{2} A_{n+1/2}}G_{n+1/2}G_{n+1/2}^{\dag}\mathfrak{e}^{\frac{\tau}{2} A_{n+1/2}^{\dag}}.
  \end{equation*}
  By the definition of $\vartheta_{n+1}$, we have
  \begin{eqnarray}\label{equ3.4}
    \|\vartheta_{n+1}\|_1=\|\tilde{\varrho}_{n+1}-\varphi_{n+1}+
    \varphi_{n+1}-\hat{\varrho}_{n+1}\|_1
    \leq\|\tilde{\varrho}_{n+1}-\varphi_{n+1}\|_1
    +\|\varphi_{n+1}-\hat{\varrho}_{n+1}\|_1.
  \end{eqnarray}
  Note from \eqref{equ2.15} and \eqref{equ2.17} that $\varphi_{n+1}=\tilde{X}_{n+1}\tilde{X}_{n+1}^{\dag}$, $\hat{\varrho}_{n+1}=\hat{X}_{n+1}\hat{X}_{n+1}^{\dag}$ and
  $\hat{X}_{n+1}= \mathcal{T}_{\varepsilon_1}(\tilde{X}_{n+1})$. It then follows that
  \begin{equation}\label{equ3.5}
    \|\varphi_{n+1}-\hat{\varrho}_{n+1}\|_1=
    \left\|\tilde{X}_{n+1}\tilde{X}_{n+1}^{\dag}-\mathcal{T}_{\varepsilon_1}(\tilde{X}_{n+1})
    \mathcal{T}_{\varepsilon_1}(\tilde{X}_{n+1})^{\dag}\right\|_1\leq \varepsilon_1.
  \end{equation}

  Note from \eqref{equ2.17a} and \eqref{equ2.7} that
  \begin{equation*}
    \tilde{\varrho}_{n+1}=e^{\tau A_{n+1/2}}\varrho_ne^{\tau A_{n+1/2}^{\dag}}
    +\tau e^{\frac{\tau}{2} A_{n+1/2}}\hat{F}(t_{n+1/2},\tilde{\varrho}_{n+1/2})e^{\frac{\tau}{2} A_{n+1/2}^{\dag}},
  \end{equation*}
  where
  \[\tilde{\varrho}_{n+1/2}=e^{\frac{\tau}{2} A_n}\left(\varrho_n+\frac{\tau}{2} \hat{F}(t_n,\varrho_n)\right)e^{\frac{\tau}{2} A_n^{\dag}},\]
  and
  \[\hat{F}(t,\rho)=\sum_{k=1}^K\gamma_k(t) \, L_k \, \rho \,L_k^{\dag}.\]
  It follows that
  \begin{eqnarray}\label{equ3.6}
    \|\tilde{\varrho}_{n+1}-\varphi_{n+1}\|_1 &\leq& \underbrace{\left\|e^{\tau A_{n+1/2}}\varrho_ne^{\tau A_{n+1/2}^{\dag}}-
    \mathfrak{e}^{\tau A_{n+1/2}}\varrho_n\mathfrak{e}^{\tau A_{n+1/2}^{\dag}}\right\|_1}_{:=I_1} \nonumber\\
     &&+ \underbrace{
    \left\|e^{\frac{\tau}{2} A_{n+1/2}}(\tau\hat{F}(t_{n+1/2},\tilde{\varrho}_{n+1/2})-\tilde{G}_{n+1/2}\tilde{G}_{n+1/2}^{\dag})e^{\frac{\tau}{2} A_{n+1/2}^{\dag}}\right\|_1}_{:=I_2} \nonumber \\
    &&+\underbrace{\left\|e^{\frac{\tau}{2} A_{n+1/2}}\tilde{G}_{n+1/2}\tilde{G}_{n+1/2}^{\dag}e^{\frac{\tau}{2} A_{n+1/2}^{\dag}}-\mathfrak{e}^{\frac{\tau}{2} A_{n+1/2}}\tilde{G}_{n+1/2}\tilde{G}_{n+1/2}^{\dag}\mathfrak{e}^{\frac{\tau}{2} A_{n+1/2}^{\dag}}\right\|_1}_{:=I_3}\nonumber\\
    &&+\underbrace{\left\|\mathfrak{e}^{\frac{\tau}{2} A_{n+1/2}}(\tilde{G}_{n+1/2}\tilde{G}_{n+1/2}^{\dag}-G_{n+1/2}G_{n+1/2}^{\dag})\mathfrak{e}^{\frac{\tau}{2} A_{n+1/2}^{\dag}}\right\|_1}_{:=I_4}.
  \end{eqnarray}

  By the assumption on the matrix exponential algorithm and note that $\|\varrho_n\|_1=1$, we obtain
\begin{equation}\label{equ3.7}
  I_1\leq C_e\varepsilon_2\|\varrho_n\|_1=C_e \, \varepsilon_2.
\end{equation}
Using the following inequality
\begin{eqnarray*}
  \|\mathfrak{e}^{\frac{\tau}{2} A_{n+1/2}}\sigma\mathfrak{e}^{\frac{\tau}{2} A_{n+1/2}^{\dag}}\|_1
  &\leq& \|\mathfrak{e}^{\frac{\tau}{2} A_{n+1/2}}\sigma\mathfrak{e}^{\frac{\tau}{2} A_{n+1/2}^{\dag}}-
  e^{\frac{\tau}{2} A_{n+1/2}}\sigma e^{\frac{\tau}{2} A_{n+1/2}^{\dag}}\|_1+\|e^{\frac{\tau}{2} A_{n+1/2}}\sigma e^{\frac{\tau}{2} A_{n+1/2}^{\dag}}\|_1\\
  &\leq& (1+C_e\varepsilon_2)\|\sigma\|_1,
\end{eqnarray*}
and noting that $G_{n+1/2}=\mathcal{T}_{\varepsilon_1}(\tilde{G}_{n+1/2})$, we obtain
\begin{equation}\label{equ3.8}
  I_4\leq (1+C_e\varepsilon_2)\left\|\tilde{G}_{n+1/2}\tilde{G}_{n+1/2}^{\dag}-G_{n+1/2}G_{n+1/2}^{\dag}\right\|_1
  \leq (1+C_e\varepsilon_2)\varepsilon_1.
\end{equation}

Note from \eqref{equ2.15} that $\tilde{G}_{n+1/2}\tilde{G}_{n+1/2}^{\dag}=\tau \hat{F}(t_{n+1/2},X_{n+1/2}X_{n+1/2}^{\dag})$ and $X_{n+1/2}=\mathcal{T}_{\varepsilon_1}(\tilde{X}_{n+1/2})$, we have
\begin{eqnarray}\label{equ3.9}
  I_2&\leq& \tau \left\|\hat{F}(t_{n+1/2},\tilde{\varrho}_{n+1/2})-\hat{F}(t_{n+1/2},X_{n+1/2}X_{n+1/2}^{\dag})\right\|_1 \nonumber \\
  &\leq& \tau C_2\left\|\tilde{\varrho}_{n+1/2}-X_{n+1/2}X_{n+1/2}^{\dag}\right\|_1\nonumber \\
  &\leq& \tau C_2\left\|\tilde{\varrho}_{n+1/2}-\tilde{X}_{n+1/2}\tilde{X}_{n+1/2}^{\dag}\right\|_1
  +\tau C_2\left\|\tilde{X}_{n+1/2}\tilde{X}_{n+1/2}^{\dag}-X_{n+1/2}X_{n+1/2}^{\dag}\right\|_1\nonumber \\
  &\leq& \tau C_2\left\|e^{\frac{\tau}{2} A_n}\left(\varrho_n+\frac{\tau}{2} \hat{F}(t_n,\varrho_n)\right)e^{\frac{\tau}{2} A_n^{\dag}}-
  \mathfrak{e}^{\frac{\tau}{2} A_n}\left(\varrho_n+\frac{1}{2} G_nG_n^{\dag}\right)\mathfrak{e}^{\frac{\tau}{2} A_n^{\dag}}\right\|_1+\tau C_2\varepsilon_1\nonumber \\
  &\leq&\frac{\tau C_2}{2}\left\|\tau e^{\frac{\tau}{2} A_n}\hat{F}(t_n,\varrho_n)e^{\frac{\tau}{2} A_n^{\dag}}-\mathfrak{e}^{\frac{\tau}{2} A_n} G_nG_n^{\dag}\mathfrak{e}^{\frac{\tau}{2} A_n^{\dag}}\right\|_1 +\tau C_2(C_e\varepsilon_2+\varepsilon_1)\nonumber \\
  &\leq& \frac{\tau C_2}{2}\left\|\tau e^{\frac{\tau}{2} A_n}\hat{F}(t_n,\varrho_n)e^{\frac{\tau}{2} A_n^{\dag}}-\tau\mathfrak{e}^{\frac{\tau}{2} A_n} \hat{F}(t_n,\varrho_n)\mathfrak{e}^{\frac{\tau}{2} A_n^{\dag}}\right\|_1 \nonumber\\
  &&+\frac{\tau C_2}{2}\left\|\mathfrak{e}^{\frac{\tau}{2} A_n} (\tilde{G}_n\tilde{G}_n^{\dag}-G_nG_n^{\dag})\mathfrak{e}^{\frac{\tau}{2} A_n^{\dag}}\right\|_1+\tau C_2(C_e\varepsilon_2+\varepsilon_1)\nonumber \\
  &\leq& \frac{1}{2}\tau^2C_2C_e\varepsilon_2\|\hat{F}(t_n,\varrho_n)\|_1
  +\frac{1}{2}\tau C_2(1+C_e\varepsilon_2)\varepsilon_1+\tau C_2(C_e\varepsilon_2+\varepsilon_1)\nonumber \\
  &\leq& \frac{1}{2}\tau C_2(3+C_e\epsilon_2)\varepsilon_1+\frac{1}{2}\tau C_2C_e(2+\tau C_2)\varepsilon_2.
\end{eqnarray}
Similarly, for the term $I_3$, we have
\begin{eqnarray}\label{equ3.10}
  I_3&\leq& \tau C_e\varepsilon_2 \left\|\hat{F}(t_{n+1/2},X_{n+1/2}X_{n+1/2}^{\dag})\right\|_1 \nonumber\\
  &\leq& \tau C_2C_e\varepsilon_2\left\|X_{n+1/2}X_{n+1/2}^{\dag}-\tilde{X}_{n+1/2}\tilde{X}_{n+1/2}^{\dag}\right\|_1
  +\tau C_2C_e\varepsilon_2\left\|\tilde{X}_{n+1/2}\tilde{X}_{n+1/2}^{\dag}\right\|_1\nonumber\\
  &\leq& \tau C_2C_e\varepsilon_2\left\|\mathfrak{e}^{\frac{\tau}{2} A_n} \left(\varrho_n+\frac{1}{2}G_nG_n^{\dag}\right)\mathfrak{e}^{\frac{\tau}{2} A_n^{\dag}}\right\|_1+\tau C_2C_e\varepsilon_2\varepsilon_1\nonumber\\
  &\leq& \tau C_2C_e\varepsilon_2(1+C_e\varepsilon_2)\left(1+\frac{1}{2}\|G_nG_n^{\dag}-\tilde{G}_n\tilde{G}_n^{\dag}\|_1
  +\frac{1}{2}\|\tilde{G}_n\tilde{G}_n^{\dag}\|_1\right)+\tau C_2C_e\varepsilon_2\varepsilon_1\nonumber\\
  &\leq& \tau C_2C_e\varepsilon_2(1+C_e\varepsilon_2)\left(1+\frac{1}{2}\varepsilon_1
  +\frac{1}{2}\tau C_2\right)+\tau C_2C_e\varepsilon_2\varepsilon_1.
\end{eqnarray}
The desired result follows from \eqref{equ3.4}-\eqref{equ3.10} with
$\tilde{c}_1=2+C_e+2TC_2C_e+TC_2(3+C_e^2)/2$ and $\tilde{c}_2=C_e+C_eTC_2(2+TC_2)(1+C_e/2)$.
\end{proof}

Now we derive the error estimate of the LREM scheme \eqref{equ2.15}.
We assume that the initial low-rank approximation satisfies
\[\|\rho_0-\varrho_0\|_1\leq\delta,\]
for some $\delta>0$.

\begin{theorem}\label{thm3.3}
 Assume that the error tolerances of the column compression and matrix exponential algorithm satisfy $\varepsilon_1=\tau\epsilon_1$ and $\varepsilon_2=\tau\epsilon_2$ for some $\epsilon_1,~\epsilon_2>0$, respectively. Let $\rho(t)$ be the solution of the Lindblad equation \eqref{equ2.2} and $\{\varrho_n\}_{n=0}^N$ be the numerical solution generated by the LREM scheme \eqref{equ2.15}. Then it holds that
  \[\|\rho(t_n)-\varrho_n\|_1\leq c_1 \, \tau^2+c_2 \, \delta+c_3 \, \epsilon_1+c_4 \, \epsilon_2,\qquad 0\leq n\leq N,\]
  where the positive constants $c_1$, $c_2$, $c_3$, $c_4$ depend on $C_1$, $C_2$, $C_e$ and $T$ but are independent of $\tau$, $n$, $\delta$, $\epsilon_1$ and $\epsilon_2$.
\end{theorem}
\begin{proof}
   We first split the global error $\rho(t_{n+1})-\varrho_{n+1}$ as follows:
\begin{equation}\label{equ3.11}
  \rho(t_{n+1})-\varrho_{n+1}=(\rho(t_{n+1})-\rho_{n+1})+(\rho_{n+1}-\check{\rho}_{n+1})+(\check{\rho}_{n+1}-\varrho_{n+1}),
\end{equation}
where the auxiliary quantities $\rho_{n+1}$ and $\check{\rho}_{n+1}$ are derived from the unnormalized FREM scheme \eqref{equ2.7} with initial value $\rho_0$ and low-rank initial value $\varrho_0$, respectively.  In other words,
  \begin{align}
    & \rho_{n+1}=\Phi(t_n,\rho_n),\qquad 0\leq n\leq N-1,\label{equ3.12}\\
    & \check{\rho}_{n+1}=\Phi(t_n,\check{\rho}_{n}),\qquad 0\leq n\leq N-1,\label{equ3.13}
  \end{align}
where $\check{\rho}_0=\varrho_0$. Note that the first component $\rho(t_{n+1})-\rho_{n+1}$ in \eqref{equ2.11} denotes the global error of the unnormalized FREM scheme \eqref{equ2.7}. We apply Theorem \ref{thm3.1} to find
\begin{equation}\label{equ3.14}
  \|\rho(t_{n+1})-\rho_{n+1}\|_1\leq C_3 \, \tau^2.
\end{equation}

The second component $\rho_{n+1}-\check{\rho}_{n+1}$ is the difference between the full-rank solutions with initial values $\rho_0$ and low-rank $\varrho_0$.
Subtracting \eqref{equ3.13} from \eqref{equ3.12} and applying Lemma \ref{lem3.2}, we obtain
\begin{eqnarray}\label{equ3.15}
  \|\rho_{n+1}-\check{\rho}_{n+1}\|_1&=&\|\Phi(t_n,\rho_n)-\Phi(t_n,\check{\rho}_n)\|_1\nonumber\\
  &\leq&(1+\tau C_2 +\tau^2C_2^2/2)\|\rho_n-\check{\rho}_n\|_1 \nonumber\\
  &\leq&(1+\tau C_2 +\tau^2C_2^2/2)^{n+1}\|\rho_0-\check{\rho}_0\|_1\leq c_2 \, \delta,
\end{eqnarray}
where $c_2=e^{C_2T}$.

The third component $\check{\rho}_{n+1}-\varrho_{n+1}$ in \eqref{equ2.11} is the difference of the solutions obtained with the FREM scheme \eqref{equ2.7} and the LREM scheme \eqref{equ2.15} with the same low-rank initial value $\varrho_0$.
By using \eqref{equ2.17}, Lemmas \ref{lem3.2}, \ref{lem3.3} and \ref{lem3.4}, we get
\begin{eqnarray}\label{equ3.16}
  \|\check{\rho}_{n+1}-\varrho_{n+1}\|_1&\leq& \|\check{\rho}_{n+1}-\tilde{\varrho}_{n+1}\|_1
  +\|\tilde{\varrho}_{n+1}-\hat{\varrho}_{n+1}\|_1+\|\hat{\varrho}_{n+1}-\varrho_{n+1}\|_1\nonumber\\
  &=&\|\Phi(t_n,\check{\rho}_n)-\Phi(t_n,\varrho_n)\|_1+\|\vartheta_{n+1}\|_1+\|(\Tr(\hat{\varrho}_{n+1})-1)\varrho_{n+1}\|_1  \nonumber\\
  &\leq& (1+\tau C_2+\tau^2C_2^2/2)\|\check{\rho}_n-\varrho_n\|_1+\|\vartheta_{n+1}\|_1+\|(\Tr(\hat{\varrho}_{n+1})-1)\varrho_{n+1}\|_1\nonumber\\
  &=&(1+\tau C_2+\tau^2C_2^2/2)\|\check{\rho}_n-\varrho_n\|_1+\|\vartheta_{n+1}\|_1+|\Tr(\tilde{\varrho}_{n+1})-\Tr(\vartheta_{n+1})-1|\nonumber\\
  &\leq&(1+\tau C_2+\tau^2C_2^2/2)\|\check{\rho}_n-\varrho_n\|_1+\|\vartheta_{n+1}\|_1+
  |\Tr(\tilde{\varrho}_{n+1})-1|+|\Tr(\vartheta_{n+1})|\nonumber\\
  &\leq&(1+\tau C_2+\tau^2C_2^2/2)\|\check{\rho}_n-\varrho_n\|_1+2\|\vartheta_{n+1}\|_1+
  |\Tr(\Phi(t_n,\varrho_n))-1|\nonumber\\
  &\leq&(1+\tau C_2+\tau^2C_2^2/2)\|\check{\rho}_n-\varrho_n\|_1+2\tau(\tilde{c}_1\epsilon_1+\tilde{c}_2\epsilon_2)+C_1\tau^3.
\end{eqnarray}
By recursion and noting that $\check{\rho}_0=\varrho_0$, we obtain
\begin{eqnarray}\label{equ3.17}
    \|\check{\rho}_{n+1}-\varrho_{n+1}\|_1&\leq&(1+\tau C_2+\tau^2C_2^2/2)^{n+1}\|\check{\rho}_0-\varrho_0\|_1\nonumber\\
    &&+(2\tilde{c}_1\tau\epsilon_1+2\tilde{c}_2\tau\epsilon_2+C_1\tau^3)\sum_{j=0}^{n}(1+\tau C_2+\tau^2C_2^2/2)^{j} \nonumber\\
    &\leq& (c_1-C_3)\tau^2 + c_3\epsilon_1+ c_4\epsilon_2,
  \end{eqnarray}
  where $c_1=C_1(e^{TC_2}-1)/C_2+C_3$, $c_3=2\tilde{c}_1(e^{TC_2}-1)/C_2$ and $c_4=2\tilde{c}_2(e^{TC_2}-1)/C_2$. Combining \eqref{equ3.11}, \eqref{equ3.14},
  \eqref{equ3.15} and \eqref{equ3.17} completes the proof.
\end{proof}

\section{Error analysis for backward problem}\label{sec-err2}
In this section, we consider the error estimates of the FREM scheme \eqref{equ2.11} and the LREM scheme \eqref{equ2.18} for the backward Lindblad equation \eqref{equ2.3}. First, we present several auxiliary results to support the error estimates.

\begin{lemma}\label{lem4.1}
 (see \cite{Chen3}) For any matrix $B\in \mathbb{C}^{m\times m}$ and any Hermitian matrix $\sigma\in \mathbb{C}^{m\times m}$, it holds that
  \[\left\|B \, \sigma \, B^{\dag}\right\|_1\leq \left\|B \, |\sigma| \, B^{\dag}\right\|_1,\]
  where $|\sigma|=\sqrt{\sigma^{\dag}\sigma}$.
\end{lemma}

\begin{lemma}\label{lem4.2}
   For any Hermitian matrix $\sigma\in \mathbb{C}^{m\times m}$, it holds for $s\in[0,T]$ that
  \begin{equation*}
    \left\|e^{tA(s)^{\dag}} \, \sigma \, e^{tA(s)}\right\|_1\leq \left\|\sigma\right\|_1,\qquad \forall t\geq0.
  \end{equation*}
\end{lemma}
\begin{proof}
  Denote $|\sigma|=\sqrt{\sigma^{\dag}\sigma}$, we see that $|\sigma|\geq0$ and $\|\sigma\|_1=\Tr(|\sigma|)$.
 For any $t\geq0$, based on Lemma \ref{lem4.1}, we have
  \begin{equation}\label{equ4.1}
    \left\|e^{tA(s)^{\dag}} \, \sigma \, e^{tA(s)}\right\|_1\leq \left\|e^{tA(s)^{\dag}} \, |\sigma| \, e^{tA(s)}\right\|_1 =\Tr\left(e^{tA(s)^{\dag}} \, |\sigma| \, e^{tA(s)}\right).
  \end{equation}
  Denote $\varrho(t)=e^{tA(s)^{\dag}} \, |\sigma| \, e^{tA(s)}$, we see that $\varrho(t)$ is the solution of the differential equation
  \begin{equation}\label{equ4.2}
    \dot{\varrho}=A(s)^{\dag} \, \varrho+\varrho \, A(s),\qquad \varrho(0)=|\sigma|.
  \end{equation}
  Recall the definition of $A(s)$ in \eqref{equ2.1}, we can rewrite  \eqref{equ4.2} as
  \begin{equation*}
    \dot{\varrho}=i H(s)\varrho-i \varrho H(s)-\frac{1}{2}\sum_{k=1}^K\gamma_k(s)\left(L_k^{\dag}L_k \, \varrho+\varrho \, L_k^{\dag}L_k\right),\qquad \varrho(0)=|\sigma|.
  \end{equation*}
  By the variation-of-constants formula, we have
  \begin{equation*}
    \varrho(t)=e^{itH(s)}|\sigma| \, e^{-itH(s)}-\frac{1}{2}\sum_{k=1}^K\gamma_k(s)\int_0^te^{i(t-v)H(s)}\left(L_k^{\dag}L_k \, \varrho(v)+\varrho(v) \, L_k^{\dag}L_k\right) e^{-i(t-v)H(s)} \, dv.
  \end{equation*}
  It then follows that
  \begin{eqnarray}\label{equ4.3}
   \Tr(\varrho(t))&=& \Tr(e^{tA(s)^{\dag}} \, |\sigma| \, e^{tA(s)})=\Tr(e^{itH(s)}|\sigma| e^{-itH(s)})\nonumber\\
    &&-\frac{1}{2}\sum_{k=1}^K\gamma_k(s)\int_0^t\Tr\left(e^{i(t-v)H(s)}\left(L_k^{\dag}L_k\varrho(v)+\varrho(v) L_k^{\dag}L_k\right) e^{-i(t-v)H(s)}\right)dv\nonumber\\
    &=&\Tr(|\sigma|)-\sum_{k=1}^K\gamma_k(s)\int_0^t\Tr\left(L_k\varrho(v)L_k^{\dag} \right)dv\nonumber\\
    &=&\|\sigma\|_1-\sum_{k=1}^K\gamma_k(s)\int_0^t\left\|L_ke^{vA(s)^{\dag}} \, |\sigma| \, e^{vA(s)}L_k^{\dag}\right\|_1dv\nonumber\\
    &\leq& \|\sigma\|_1.
  \end{eqnarray}
  Combining \eqref{equ4.1} and \eqref{equ4.3} completes the proof.
\end{proof}

\subsection{Error estimate of the FREM scheme}
We first consider the consistency of the FREM scheme \eqref{equ2.11}. Using error estimates of the basic right-rectangle quadrature formula and midpoint quadrature formula, it then follows from \eqref{equ2.10} that
\begin{subequations}\label{equ4.4}
\begin{align}
  &q(t_{n+1/2})=e^{\frac{\tau}{2} A_{n+1}^{\dag}} \left( q(t_{n+1}) + \frac{\tau}{2}\sum_{k=1}^K\gamma_k(t_{n+1}) \, L_k^{\dag} \, q(t_{n+1}) \,L_k \right) e^{\frac{\tau}{2} A_{n+1}}+\mathcal{O}(\tau^2) ,\label{equ4.4a}\\
  &q(t_{n})=e^{\tau A_{n+1/2}^{\dag}} \, q(t_{n+1}) \, e^{\tau A_{n+1/2}} +
  \tau \sum_{k=1}^K\gamma_k(t_{n+1/2}) \,e^{\frac{\tau}{2} A_{n+1/2}^{\dag}} \, L_k^{\dag} \, q(t_{n+1/2}) \,L_k \, e^{\frac{\tau}{2} A_{n+1/2}}+\mathcal{O}(\tau^3).\label{equ4.4b}
\end{align}
\end{subequations}
Inserting \eqref{equ4.4a} into \eqref{equ4.4b} yields
\begin{equation}\label{equ4.5}
  q(t_{n})=\Psi(t_{n+1},q(t_{n+1}))+\tilde{R}_n,
\end{equation}
and the truncation error $\tilde{R}_n$ satisfies
\begin{equation}\label{equ4.6}
  \max_{0\leq n\leq N-1}\|\tilde{R}_n\|_1\leq \tilde{C}_1 \, \tau^3,
\end{equation}
where the positive constant $\tilde{C}_1$ depends on $L_k$, $A(t)$, $\gamma_k(t)$, $q(t)$ and their first and second order derivatives.

\begin{lemma}\label{lem4.3}
 For any Hermitian matrices $q_{n+1},p_{n+1}\in \mathbb{C}^{m\times m}$, it holds that
 \[ \|\Psi(t_{n+1},q_{n+1})-\Psi(t_{n+1},p_{n+1})\|_1\leq (1+\tau C_2+\tau^2C_2^2/2) \, \|q_{n+1}-p_{n+1}\|_1,\]
 $n=N-1,\ldots,0$.
\end{lemma}
\begin{proof}
  Considering the difference of the equations \eqref{equ2.11a} with respect to different initial values $q_{n+1}$ and $p_{n+1}$ and using Lemma \ref{lem4.2}, we get
  \begin{eqnarray*}
    \|q_{n+1/2}-p_{n+1/2}\|_1&\leq & \|q_{n+1}-p_{n+1}\|_1+\frac{\tau }{2} \sum_{k=1}^{K}\gamma_k(t_{n+1})\|L_k^{\dag}(q_{n+1}-p_{n+1})L_k\|_1\\
    &\leq& (1+\frac{\tau}{2}C_2)\|q_{n+1}-p_{n+1}\|_1.
  \end{eqnarray*}
  Similarly, using \eqref{equ2.11b}, Lemma \ref{lem4.2}, $q_{n}=\Psi(t_{n+1},q_{n+1})$ and $p_{n}=\Psi(t_{n+1},p_{n+1})$, we have
  \begin{eqnarray*}
  \|q_{n}-p_{n}\|_1 &=&\|\Psi(t_{n+1},q_{n+1})-\Psi(t_{n+1},p_{n+1})\|_1\\
   &\leq& \|q_{n+1}-p_{n+1}\|_1+\tau \sum_{k=1}^{K}\gamma_k(t_{n+1/2})\|L_k^{\dag}(q_{n+1/2}-p_{n+1/2})L_k\|_1\\
    &\leq&\|q_{n+1}-p_{n+1}\|_1+\tau C_2 \|q_{n+1/2}-p_{n+1/2}\|_1\\
    &=&(1+\tau C_2+\tau^2 C_2^2/2)\|q_{n+1}-p_{n+1}\|_1,
  \end{eqnarray*}
  which completes the proof.
\end{proof}

\begin{lemma}\label{lem4.4}
  Let $p$ be Hermitian and positive semidefinite with unit trace, then it holds that
  \[1-\tilde{C}_1 \, \tau^3\leq\Tr(\Psi(t_{n+1},p))\leq 1+\tilde{C}_1 \, \tau^3,\qquad n=N-1,\ldots,0.\]
\end{lemma}
\begin{proof}
  Denote by $g(t)$ the solution of the adjoint Lindblad equation \eqref{equ2.3} with terminal condition $g(t_{n+1})=p$. Then, we have that $g(t)$ is Hermitian and positive semidefinite with unit trace, i.e., $\|g(t)\|_1=\Tr(g(t))=1$ for all $0\leq t\leq t_{n+1}$. Note that
  $\Psi(t_{n+1},p)$ is the numerical approximation to $g(t_{n})$ by using the FREM scheme \eqref{equ2.11} for a single step with exact terminal value $g(t_{n+1})$.
  Therefore, the consistency \eqref{equ4.5}-\eqref{equ4.6} of the FREM scheme \eqref{equ2.11} gives
  \[\|\Psi(t_{n+1},p)-g(t_{n})\|_1\leq \tilde{C}_1 \, \tau^3.\]
 Note that
 \[\|g(t_{n})\|_1-\|\Psi(t_{n+1},p)-g(t_{n})\|_1\leq\|\Psi(t_{n+1},p)\|_1\leq
 \|g(t_{n})\|_1+\|\Psi(t_{n+1},p)-g(t_{n})\|_1.\]
 The statement now follows from the fact that the FREM scheme \eqref{equ2.11} is positivity preserving and $\|g(t_{n})\|_1=1$.
\end{proof}

The following result shows the second-order convergence of the unnormalized FREM scheme \eqref{equ2.11}.
\begin{theorem}\label{thm4.1}
  The numerical solution $q_n$ generated by the unnormalized FREM scheme \eqref{equ2.11} with $q_N=Q$ satisfies the error estimate
  \[\|q(t_n)-q_n\|_1\leq \tilde{C}_3 \, \tau^2,\qquad 0\leq n\leq N,\]
  where the constant $\tilde{C}_3>0$ depends on $\tilde{C}_1,~C_2,~T$ but is independent of $\tau$ and $n$.
\end{theorem}
\begin{proof}
  Subtracting $q_n=\Psi(t_{n+1},q_{n+1})$ from \eqref{equ4.5}, applying \eqref{equ4.6} and Lemma \ref{lem4.3} yields
  \begin{eqnarray*}
    \|q(t_n)-q_n\|_1&\leq&\|\Psi(t_{n+1},q(t_{n+1}))-\Psi(t_{n+1},q_{n+1})\|_1+\|\tilde{R}_n\|_1\\
    &\leq& (1+\tau C_2+\tau^2C_2^2/2)\|q(t_{n+1})-q_{n+1}\|_1 + \tilde{C}_1\tau^3.
  \end{eqnarray*}
  Then we have
  \begin{equation*}
    \|q(t_n)-q_n\|_1\leq (1+\tau C_2+\tau^2C_2^2/2)^{N-n}\|q(T)-q_N\|_1
    +\tilde{C}_1\tau^3\sum_{j=0}^{N-n-1}(1+\tau C_2+\tau^2C_2^2/2)^j.
  \end{equation*}
  Noting that $q_N=q(T)$, so that the statement holds with $\tilde{C}_3=(e^{C_2T}-1)\tilde{C}_1/C_2$.
\end{proof}

The convergence of the normalized FREM scheme \eqref{equ2.12b} for the backward Lindblad equation is stated in the following result.

\begin{theorem}\label{thm4.2}
    The numerical solution $q_n$ generated by the normalized FREM scheme \eqref{equ2.12b} with $q_N=Q$ satisfies the error estimate
  \[\|q(t_n)-q_n\|_1\leq 2\tilde{C}_3 \, \tau^2,\qquad 0\leq n\leq N,\]
  where the constant $\tilde{C}_3$ is as defined in Theorem \ref{thm4.1}.
  \end{theorem}
  \begin{proof}
    Let $\hat{q}_n$ be the numerical solution derived from the unnormalized FREM scheme \eqref{equ2.11}, i.e., $\hat{q}_n=\Psi(t_{n+1},\hat{q}_{n+1})$ with $\hat{q}_N=Q$. From Theorem \ref{thm4.1} we then get
    \begin{eqnarray*}
      \|q(t_n)-q_n\|_1\leq \|q(t_n)-\hat{q}_n\|_1+\|\hat{q}_n-q_n\|_1
      \leq \tilde{C}_3\tau^2 + \|\hat{q}_n-q_n\|_1.
    \end{eqnarray*}
    By the triangle inequality we get
    \begin{eqnarray*}
      \|\hat{q}_n-q_n\|_1 &\leq& \|\hat{q}_n-\tilde{q}_n\|_1+\|\tilde{q}_n-q_n\|_1 \\
      &=&\|\Psi(t_{n+1},\hat{q}_{n+1})-\Psi(t_{n+1},q_{n+1})\|_1 +\|(\Tr(\tilde{q}_n)-1)q_n\|_1\\
      &=&\|\Psi(t_{n+1},\hat{q}_{n+1})-\Psi(t_{n+1},q_{n+1})\|_1 + |\Tr(\Psi(t_{n+1},q_{n+1}))-1|\cdot\|q_n\|_1.
    \end{eqnarray*}
    Further it follows from Lemmas \ref{lem4.3} and \ref{lem4.4} and from $q_n\geq0$ with $\|q_n\|_1=1$ that
    \begin{eqnarray*}
      \|\hat{q}_n-q_n\|_1 &\leq& (1+\tau C_2+\tau^2C_2^2/2)\|\hat{q}_{n+1}-q_{n+1}\|_1+\tilde{C}_1\tau^3\\
      &\leq& (1+\tau C_2+\tau^2C_2^2/2)^{N-n}\|\hat{q}_{N}-q_{N}\|_1+\tilde{C}_3\tau^2.
    \end{eqnarray*}
    The desired estimate then follows from $\hat{q}_{N}=q_{N}=Q$.
  \end{proof}

\subsection{Error estimate of the LREM scheme}
Our next aim is to estimate the error of the LREM scheme \eqref{equ2.18} for the backward Lindblad equation. First, we present the following result concerning the bound of $\theta_n$ as defined in \eqref{equ2.20}.

\begin{lemma}\label{lem4.5}
 Let $\varepsilon_1>0$ be the error tolerance of the column compression algorithm used in the LREM scheme \eqref{equ2.18}. Assume that the matrix exponential algorithm used in the LREM scheme \eqref{equ2.18} satisfies $\|e^{\mu\tau A_{\nu}^{\dag}}\sigma e^{\mu\tau A_{\nu}}-\mathfrak{e}^{\mu\tau A_{\nu}^{\dag}}\sigma \mathfrak{e}^{\mu\tau A_{\nu}}\|_1\leq \tilde{C}_e\varepsilon_2\|\sigma\|_1$ for $\mu=0.5, 1$, $\nu=n,n+1/2$ with $n=0,\ldots,N-1$ and any $\sigma\in \mathbb{C}^{m\times m}$, where $0<\varepsilon_2<1$ is the
   corresponding error tolerance and $\tilde{C}_e>0$ is the error constant. Then it holds that
  \begin{equation*}
  \|\theta_{n}\|_1\leq \hat{c}_1 \, \varepsilon_1 + \hat{c}_2  \, \varepsilon_2,\qquad n=0,1,\ldots,N-1,
\end{equation*}
where the positive constants $\hat{c}_1,~\hat{c}_2$ depend on $\tilde{C}_e$, $C_2$ and $T$ but is independent of $\tau$, $n$, $\varepsilon_1$ and $\varepsilon_2$.
\end{lemma}
\begin{proof}
  With the notation
  \begin{equation*}
    \phi_{n}=\mathfrak{e}^{\tau A_{n+1/2}^{\dag}}p_{n+1}\mathfrak{e}^{\tau A_{n+1/2}}+
    \mathfrak{e}^{\frac{\tau}{2} A_{n+1/2}^{\dag}}W_{n+1/2}W_{n+1/2}^{\dag}\mathfrak{e}^{\frac{\tau}{2} A_{n+1/2}},
  \end{equation*}
  and the triangle inequality, we have
  \begin{eqnarray}\label{equ4.7}
    \|\theta_{n}\|_1=\|\tilde{p}_{n}-\phi_{n}+
    \phi_{n}-\hat{p}_{n}\|_1
    \leq\|\tilde{p}_{n}-\phi_{n}\|_1
    +\|\phi_{n}-\hat{p}_{n}\|_1.
  \end{eqnarray}
  Note from \eqref{equ2.18} and \eqref{equ2.20} that $\phi_n=\tilde{Y}_n\tilde{Y}_n^{\dag}$,
  $\hat{p}_n=\hat{Y}_n\hat{Y}_n^{\dag}$ and $\hat{Y}_n=\mathcal{T}_{\varepsilon_1}(\tilde{Y}_n)$. We then have
  \begin{equation}\label{equ4.8}
    \|\phi_{n}-\hat{p}_{n}\|_1=\left\|\tilde{Y}_n\tilde{Y}_n^{\dag}-\hat{Y}_n\hat{Y}_n^{\dag}\right\|_1
    \leq \varepsilon_1.
  \end{equation}

  Since $\tilde{p}_n=\Psi(t_{n+1},p_{n+1})$, straightforward calculation shows that
  \begin{equation*}
    \tilde{p}_n=e^{\tau A_{n+1/2}^{\dag}}p_{n+1}e^{\tau A_{n+1/2}}
    +\tau e^{\frac{\tau}{2} A_{n+1/2}^{\dag}}\check{F}(t_{n+1/2},\tilde{p}_{n+1/2})e^{\frac{\tau}{2} A_{n+1/2}},
  \end{equation*}
  where
  \[\tilde{p}_{n+1/2}=e^{\frac{\tau}{2} A_{n+1}^{\dag}}\left(p_{n+1}+\frac{\tau}{2} \check{F}(t_{n+1},p_{n+1})\right)e^{\frac{\tau}{2} A_{n+1}},\]
  and
  \[\check{F}(t,p)=\sum_{k=1}^K\gamma_k(t) \, L_k^{\dag} \, p \,L_k.\]
  By the triangle inequality we get
  \begin{eqnarray}\label{equ4.9}
    \|\tilde{p}_{n}-\phi_{n}\|_1 &\leq& \underbrace{\left\|e^{\tau A_{n+1/2}^{\dag}}p_{n+1}e^{\tau A_{n+1/2}}-
    \mathfrak{e}^{\tau A_{n+1/2}^{\dag}}p_{n+1}\mathfrak{e}^{\tau A_{n+1/2}}\right\|_1}_{:=I_1} \nonumber\\
     &&+ \underbrace{
    \left\|e^{\frac{\tau}{2} A_{n+1/2}^{\dag}}(\tau\check{F}(t_{n+1/2},\tilde{p}_{n+1/2})-\tilde{W}_{n+1/2}\tilde{W}_{n+1/2}^{\dag})e^{\frac{\tau}{2} A_{n+1/2}}\right\|_1}_{:=I_2} \nonumber \\
    &&+\underbrace{\left\|e^{\frac{\tau}{2} A_{n+1/2}^{\dag}}\tilde{W}_{n+1/2}\tilde{W}_{n+1/2}^{\dag}e^{\frac{\tau}{2} A_{n+1/2}}-\mathfrak{e}^{\frac{\tau}{2} A_{n+1/2}^{\dag}}\tilde{W}_{n+1/2}\tilde{W}_{n+1/2}^{\dag}\mathfrak{e}^{\frac{\tau}{2} A_{n+1/2}}\right\|_1}_{:=I_3}\nonumber\\
    &&+\underbrace{\left\|\mathfrak{e}^{\frac{\tau}{2} A_{n+1/2}^{\dag}}(\tilde{W}_{n+1/2}\tilde{W}_{n+1/2}^{\dag}-W_{n+1/2}W_{n+1/2}^{\dag})\mathfrak{e}^{\frac{\tau}{2} A_{n+1/2}}\right\|_1}_{:=I_4}.
  \end{eqnarray}

   By the assumption on the matrix exponential algorithm and note that $\|p_{n+1}\|_1=1$, we obtain
\begin{equation}\label{equ4.10}
  I_1\leq \tilde{C}_e\varepsilon_2\|p_{n+1}\|_1=\tilde{C}_e \, \varepsilon_2.
\end{equation}
Note that
\begin{eqnarray*}
  \|\mathfrak{e}^{\frac{\tau}{2} A_{n+1/2}^{\dag}}\sigma\mathfrak{e}^{\frac{\tau}{2} A_{n+1/2}}\|_1
  &\leq& \|\mathfrak{e}^{\frac{\tau}{2} A_{n+1/2}^{\dag}}\sigma\mathfrak{e}^{\frac{\tau}{2} A_{n+1/2}}-
  e^{\frac{\tau}{2} A_{n+1/2}^{\dag}}\sigma e^{\frac{\tau}{2} A_{n+1/2}}\|_1+\|e^{\frac{\tau}{2} A_{n+1/2}^{\dag}}\sigma e^{\frac{\tau}{2} A_{n+1/2}}\|_1\\
  &\leq& (1+\tilde{C}_e\varepsilon_2)\|\sigma\|_1,
\end{eqnarray*}
this combines with $W_{n+1/2}=\mathcal{T}_{\varepsilon_1}(\tilde{W}_{n+1/2})$ gives
\begin{equation}\label{equ4.11}
  I_4\leq (1+\tilde{C}_e\varepsilon_2)\left\|\tilde{W}_{n+1/2}\tilde{W}_{n+1/2}^{\dag}-W_{n+1/2}W_{n+1/2}^{\dag}\right\|_1
  \leq (1+\tilde{C}_e\varepsilon_2)\varepsilon_1.
\end{equation}

We see from \eqref{equ2.18} that $\tilde{W}_{n+1/2}\tilde{W}_{n+1/2}^{\dag}=\tau \check{F}(t_{n+1/2},Y_{n+1/2}Y_{n+1/2}^{\dag})$ and $Y_{n+1/2}=\mathcal{T}_{\varepsilon_1}(\tilde{Y}_{n+1/2})$, then we have
\begin{eqnarray}\label{equ4.12}
  I_2&\leq& \tau \left\|\check{F}(t_{n+1/2},\tilde{p}_{n+1/2})-\check{F}(t_{n+1/2},Y_{n+1/2}Y_{n+1/2}^{\dag})\right\|_1 \nonumber \\
  &\leq& \tau C_2\left\|\tilde{p}_{n+1/2}-Y_{n+1/2}Y_{n+1/2}^{\dag}\right\|_1\nonumber \\
  &\leq& \tau C_2\left\|\tilde{p}_{n+1/2}-\tilde{Y}_{n+1/2}\tilde{Y}_{n+1/2}^{\dag}\right\|_1
  +\tau C_2\left\|\tilde{Y}_{n+1/2}\tilde{Y}_{n+1/2}^{\dag}-Y_{n+1/2}Y_{n+1/2}^{\dag}\right\|_1\nonumber \\
  &\leq& \tau C_2\left\|e^{\frac{\tau}{2} A_{n+1}^{\dag}}(p_{n+1}+\frac{\tau}{2} \check{F}(t_{n+1},p_{n+1}))e^{\frac{\tau}{2} A_{n+1}}-
  \mathfrak{e}^{\frac{\tau}{2} A_{n+1}^{\dag}}(p_{n+1}+\frac{1}{2} W_{n+1}W_{n+1}^{\dag})\mathfrak{e}^{\frac{\tau}{2} A_{n+1}}\right\|_1\nonumber\\
  &&+\tau C_2\varepsilon_1\nonumber \\
  &\leq&\frac{\tau C_2}{2}\left\|\tau e^{\frac{\tau}{2} A_{n+1}^{\dag}}\check{F}(t_{n+1},p_{n+1})e^{\frac{\tau}{2} A_{n+1}}-\mathfrak{e}^{\frac{\tau}{2} A_{n+1}^{\dag}} W_{n+1}W_{n+1}^{\dag}\mathfrak{e}^{\frac{\tau}{2} A_{n+1}}\right\|_1 +\tau C_2(\tilde{C}_e\varepsilon_2+\varepsilon_1)\nonumber \\
  &\leq& \frac{\tau C_2}{2}\left\|\tau e^{\frac{\tau}{2} A_{n+1}^{\dag}}\check{F}(t_{n+1},p_{n+1})e^{\frac{\tau}{2} A_{n+1}}-\tau\mathfrak{e}^{\frac{\tau}{2} A_{n+1}^{\dag}} \check{F}(t_{n+1},p_{n+1})\mathfrak{e}^{\frac{\tau}{2} A_{n+1}}\right\|_1 \nonumber\\
  &&+\frac{\tau C_2}{2}\left\|\mathfrak{e}^{\frac{\tau}{2} A_{n+1}^{\dag}} (\tilde{W}_{n+1}\tilde{W}_{n+1}^{\dag}-W_{n+1}W_{n+1}^{\dag})\mathfrak{e}^{\frac{\tau}{2} A_{n+1}}\right\|_1+\tau C_2(\tilde{C}_e\varepsilon_2+\varepsilon_1)\nonumber \\
  &\leq& \frac{1}{2}\tau^2C_2\tilde{C}_e\varepsilon_2\|\check{F}(t_{n+1},p_{n+1})\|_1
  +\frac{1}{2}\tau C_2(1+\tilde{C}_e\varepsilon_2)\varepsilon_1+\tau C_2(\tilde{C}_e\varepsilon_2+\varepsilon_1)\nonumber \\
  &\leq& \frac{1}{2}\tau C_2(3+\tilde{C}_e\epsilon_2)\varepsilon_1+\frac{1}{2}\tau C_2\tilde{C}_e(2+\tau C_2)\varepsilon_2.
\end{eqnarray}

Similar calculations give
\begin{eqnarray}\label{equ4.13}
  I_3&\leq& \tau \tilde{C}_e\varepsilon_2 \left\|\check{F}(t_{n+1/2},Y_{n+1/2}Y_{n+1/2}^{\dag})\right\|_1 \nonumber\\
  &\leq& \tau C_2\tilde{C}_e\varepsilon_2\left\|Y_{n+1/2}Y_{n+1/2}^{\dag}-\tilde{Y}_{n+1/2}\tilde{Y}_{n+1/2}^{\dag}\right\|_1
  +\tau C_2\tilde{C}_e\varepsilon_2\left\|\tilde{Y}_{n+1/2}\tilde{Y}_{n+1/2}^{\dag}\right\|_1\nonumber\\
  &\leq& \tau C_2\tilde{C}_e\varepsilon_2\left\|\mathfrak{e}^{\frac{\tau}{2} A_{n+1}^{\dag}} \left(p_{n+1}+\frac{1}{2}W_{n+1}W_{n+1}^{\dag}\right)\mathfrak{e}^{\frac{\tau}{2} A_{n+1}}\right\|_1+\tau C_2\tilde{C}_e\varepsilon_2\varepsilon_1\nonumber\\
  &\leq& \tau C_2\tilde{C}_e\varepsilon_2(1+\tilde{C}_e\varepsilon_2)\left(1+\frac{1}{2}\|W_{n+1}W_{n+1}^{\dag}-\tilde{W}_{n+1}\tilde{W}_{n+1}^{\dag}\|_1
  +\frac{1}{2}\|\tilde{W}_{n+1}\tilde{W}_{n+1}^{\dag}\|_1\right) \nonumber \\
  &&+\tau C_2\tilde{C}_e\varepsilon_2\varepsilon_1\nonumber\\
  &\leq& \tau C_2\tilde{C}_e\varepsilon_2(1+\tilde{C}_e\varepsilon_2)\left(1+\frac{1}{2}\varepsilon_1
  +\frac{1}{2}\tau C_2\right)+\tau C_2\tilde{C}_e\varepsilon_2\varepsilon_1.
\end{eqnarray}
The desired result follows from \eqref{equ4.7}-\eqref{equ4.13} with
$\hat{c}_1=2+\tilde{C}_e+2TC_2\tilde{C}_e+TC_2(3+\tilde{C}_e^2)/2$ and $\hat{c}_2=\tilde{C}_e+\tilde{C}_eTC_2(2+TC_2)(1+\tilde{C}_e/2)$.
\end{proof}

Finally, we prove a convergence result for the LREM scheme \eqref{equ2.18}.
We assume that the terminal low-rank approximation satisfies
\[\|p_N-Q\|_1\leq\delta,\]
for some $\delta>0$.

\begin{theorem}\label{thm4.3}
 Assume that the error tolerances of the column compression and matrix exponential algorithm satisfy $\varepsilon_1=\tau\epsilon_1$ and $\varepsilon_2=\tau\epsilon_2$ for some $\epsilon_1,~\epsilon_2>0$, respectively. Let $q(t)$ be the solution of the adjoint Lindblad equation \eqref{equ2.3} and $\{p_n\}_{n=0}^N$ be the numerical solution generated by the LREM scheme \eqref{equ2.18}. Then it holds that
  \[\|q(t_n)-p_n\|_1\leq \check{c}_1 \, \tau^2+\check{c}_2 \, \delta+\check{c}_3 \, \epsilon_1+\check{c}_4 \, \epsilon_2,\qquad 0\leq n\leq N,\]
  where the positive constants $\check{c}_1$, $\check{c}_2$, $\check{c}_3$, $\check{c}_4$ depend on $\tilde{C}_1$, $C_2$, $\tilde{C}_e$ and $T$ but are independent of $\tau$, $n$, $\delta$, $\epsilon_1$ and $\epsilon_2$.
\end{theorem}
\begin{proof}
  We split the global error $q(t_{n})-p_{n}$ as follows:
\begin{equation}\label{equ4.14}
  q(t_{n})-p_{n}=(q(t_{n})-q_{n})+(q_{n}-\check{q}_{n})+(\check{q}_{n}-p_{n}),
\end{equation}
where the auxiliary quantities $q_{n}$ and $\check{q}_{n}$ are obtained from the unnormalized FREM scheme \eqref{equ2.11} with terminal value $Q$ and low-rank terminal value $p_N$, respectively.  In other words,
  \begin{align}
    & q_{n}=\Psi(t_{n+1},q_{n+1}),\qquad n=N-1,\ldots,0, \label{equ4.15}\\
    & \check{q}_{n}=\Psi(t_{n+1},\check{q}_{n+1}),\qquad n=N-1,\ldots,0,\label{equ4.16}
  \end{align}
where $q_N=Q$ and $\check{q}_N=p_N$. Note that the first component $q(t_{n})-q_{n}$ in \eqref{equ4.14} denotes the global error of the unnormalized FREM scheme \eqref{equ2.11}. Therefore, applying Theorem \ref{thm4.1} yeilds
\begin{equation}\label{equ4.17}
  \|q(t_{n})-q_{n}\|_1\leq \tilde{C}_3 \, \tau^2.
\end{equation}

The second component $q_{n}-\check{q}_{n}$ in \eqref{equ4.14} is the difference between the full-rank solutions with terminal values $Q$ and low-rank $p_N$.
Subtracting \eqref{equ4.16} from \eqref{equ4.15} and applying Lemma \ref{lem4.3} gives
\begin{eqnarray}\label{equ4.18}
  \|q_{n}-\check{q}_{n}\|_1&=&\|\Psi(t_{n+1},q_{n+1})-\Psi(t_{n+1},\check{q}_{n+1})\|_1\nonumber\\
  &\leq&(1+\tau C_2 +\tau^2C_2^2/2)\|q_{n+1}-\check{q}_{n+1}\|_1 \nonumber\\
  &\leq&(1+\tau C_2 +\tau^2C_2^2/2)^{N-n}\|q_N-\check{q}_N\|_1\leq \check{c}_2 \, \delta,
\end{eqnarray}
where $\check{c}_2=e^{C_2T}$.

The third component $\check{q}_{n}-p_{n}$ in \eqref{equ4.14} is the difference of the solutions obtained with the FREM scheme \eqref{equ2.11} and the LREM scheme \eqref{equ2.18} with the same low-rank terminal value $p_N$.
By the triangle inequality we get
\begin{eqnarray}\label{equ4.19}
  \|\check{q}_{n}-p_{n}\|_1&\leq& \|\check{q}_{n}-\tilde{p}_{n}\|_1
  +\|\tilde{p}_{n}-\hat{p}_{n}\|_1+\|\hat{p}_{n}-p_{n}\|_1,
\end{eqnarray}
where $\tilde{p}_{n}$ and  $\hat{p}_{n}$ are as defined in \eqref{equ2.20}.
It follows from \eqref{equ2.20}, Lemma \ref{lem4.4} and the fact that $p_n\geq0$ with $\|p_n\|_1=1$ that
\begin{eqnarray}\label{equ4.20}
  \|\hat{p}_{n}-p_{n}\|_1 &=& \|(\Tr(\hat{p}_n)-1)p_n\|
  =|\Tr(\tilde{p}_n)-\Tr(\theta_n)-1| \nonumber\\
  &\leq& |\Tr(\Psi(t_{n+1},p_{n+1}))-1| + |\Tr(\theta_n)| \nonumber\\
  &\leq& \|\theta_n\|_1+\tilde{C}_1\tau^3.
\end{eqnarray}
 Combining \eqref{equ4.19} and \eqref{equ4.20} and applying Lemmas \ref{lem4.4} and \ref{lem4.5}, we obtain
 \begin{eqnarray}\label{equ4.21}
   \|\check{q}_{n}-p_{n}\|_1&\leq& \|\Psi(t_{n+1},\check{q}_{n+1})-\Psi(t_{n+1},p_{n+1})\|_1
   +2\|\theta_n\|_1+\tilde{C}_1\tau^3 \nonumber\\
   &\leq& (1+\tau C_2+\tau^2C_2^2/2)\|\check{q}_{n+1}-p_{n+1}\|_1+2\tau(\hat{c}_1\epsilon_1
   +\hat{c}_2\epsilon_2)+\tilde{C}_1\tau^3 \nonumber\\
   &\leq&(1+\tau C_2+\tau^2C_2^2/2)^{N-n}\|\check{q}_{N}-p_{N}\|_1+(\check{c}_1-C_3)\tau^2+ \check{c}_3\epsilon_1\check{+c}_4\epsilon_2\nonumber\\
   &=&(\check{c}_1-\tilde{C}_3)\tau^2+ \check{c}_3\epsilon_1+\check{c}_4\epsilon_2,
 \end{eqnarray}
 where $\check{c}_1=\tilde{C}_1(e^{TC_2}-1)/C_2+\tilde{C}_3$, $\check{c}_3=2\hat{c}_1(e^{TC_2}-1)/C_2$ and $\check{c}_4=2\hat{c}_2(e^{TC_2}-1)/C_2$. By combining \eqref{equ4.14}, \eqref{equ4.17}, \eqref{equ4.18} and \eqref{equ4.21} we complete the proof.
\end{proof}

\section{Numerical experiments}\label{sec-experiments}
This section presents numerical results that validate properties of the proposed integrators. Numerical experiments presented here are implemented in Python 3.12.4 on 
a laptop with Intel(R) Core(TM) i7-8565U CPU@1.80GHz and 16GB RAM. The matrix exponential 
codes in the Python package \emph{scipy} are used in our integrators.

We use the Lindblad equation with the X-X Ising Chain Hamiltonian, as done in \cite{Chen3}, as our model for tests:
\begin{equation*}
  H(t)=\sum_{k=1}^K\left(aJ_z^{(k)}+b(J_z^{(k)})^2\right)+u(t)\sum_{k=1}^{K-1}\sum_{l=k+1}^KJ_x^{(k)}J_x^{(l)},
\end{equation*}
where
\[J_w^{(k)}=\underbrace{I_d\otimes\cdots\otimes I_d}_{k-1}\otimes J_w\otimes \underbrace{I_d\otimes\cdots\otimes I_d}_{K-k},~~~w=x,z,~~~k=1,\ldots,K,\]
$I_d$ is the $d\times d$ identity matrix and the $J_x,J_z\in \mathbb{R}^{d\times d}$ are angular momentum operators. We set $L_k=J_z^{(k)}$ and $\gamma_k(t)\equiv \gamma$.

Based on the error estimates, we display the forward state and backward state approximation errors defined as 
 \[e_{\rho}=\|\rho_N-\rho(T)\|_1,\quad e_{q}=\|q_0-q(0)\|_1\]
 for full-rank schemes and 
 \[\check{e}_{\rho}=\|X_NX_N^{\dag}-\rho(T)\|_1,\quad \check{e}_{q}=\|Y_0Y_0^{\dag}-q(0)\|_1\]
 for low-rank schemes, and then estimate the experimental order of convergence. 
The reference solutions  $\rho(T)$ and $q(0)$ are computed by the solver \emph{mesolve} developed in QuTip \cite{Johansson}.

\begin{figure}
{ \centering
\includegraphics[width=0.48\textwidth]{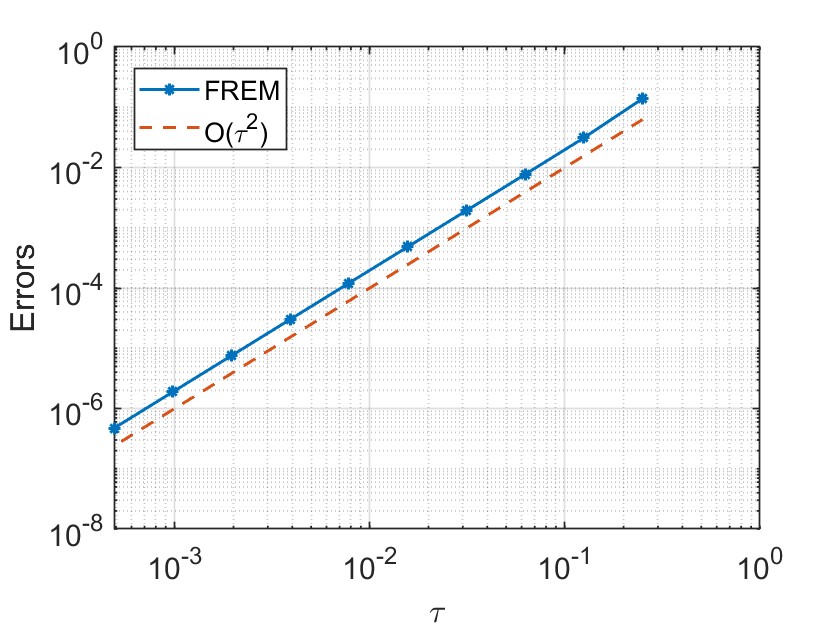}
}  { \centering
\includegraphics[width=0.48\textwidth]{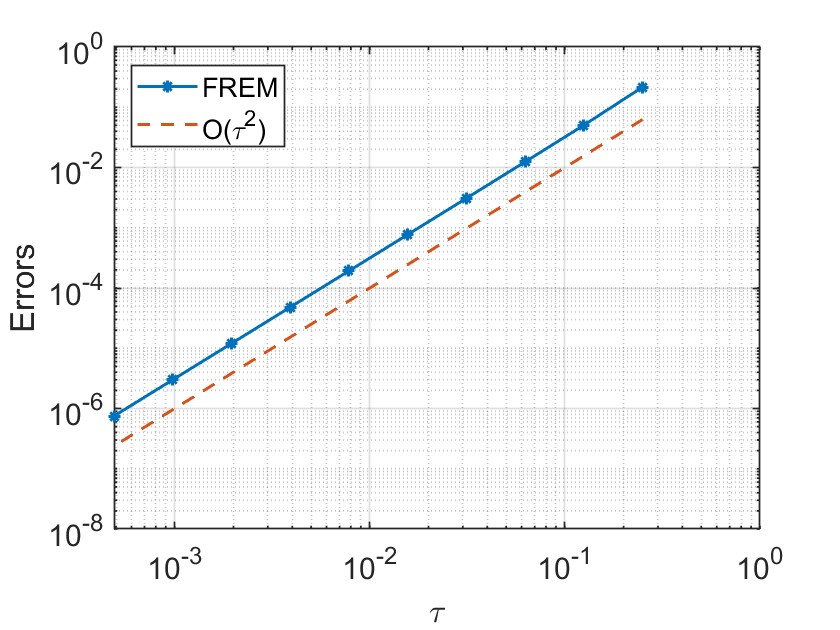}
} \caption{Numerical results of FREM schemes for the Lindblad equations with $d=6$, $K=2$, $a=1.5$, $b=1$, $\gamma=0.05$, $T=1$, $u(t)=\sin(2\pi t)$. Left: errors vs step sizes for the forward Lindblad equation. Right: errors vs step sizes for the backward Lindblad equation.}
\label{fig6.1}
\end{figure}

\begin{figure}
{ \centering
\includegraphics[width=0.48\textwidth]{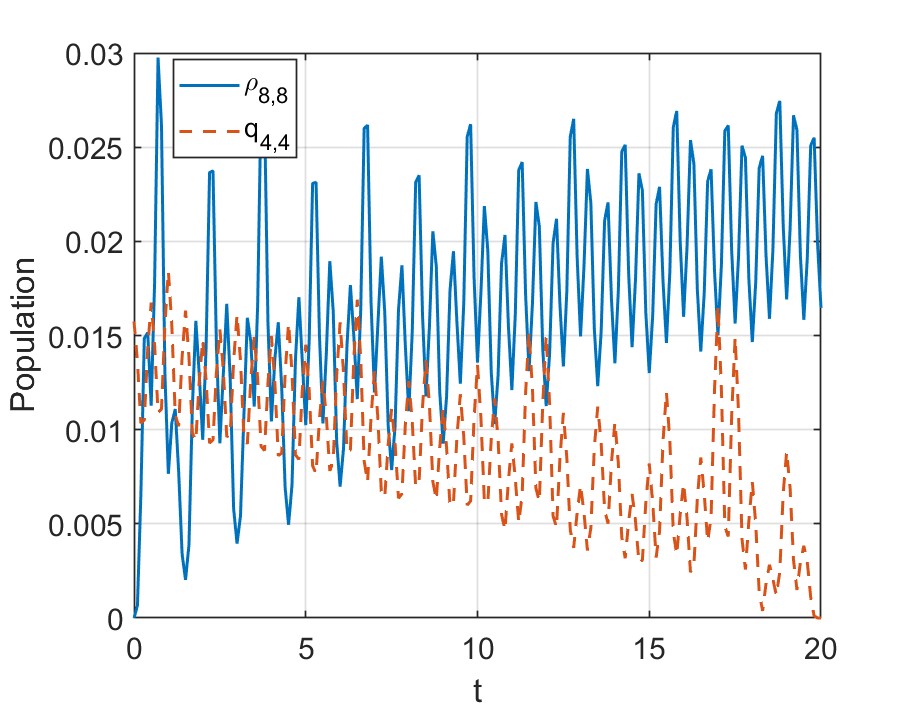}
}  { \centering
\includegraphics[width=0.48\textwidth]{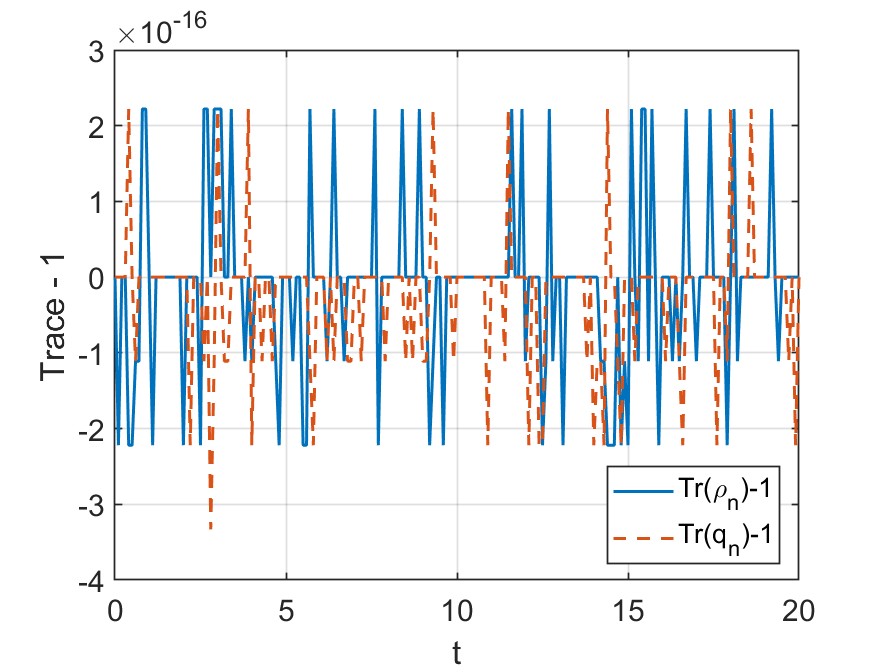}
} \caption{Numerical results of FREM schemes for the Lindblad equations with $d=6$, $K=2$, $a=1.5$, $b=1$, $\gamma=0.05$, $T=20$, $u(t)=\sin(2\pi t)$. Left: evolutions of the populations $\rho_{8,8}$ and $q_{4,4}$ with $\tau=0.1$. Right: evolutions of $\Tr(\rho_n)-1$ and $\Tr(q_n)-1$.}
\label{fig6.2}
\end{figure}

\emph{Performance of the FREM schemes.} We first investigate the convergence and structure-preserving behavior of the FREM schemes \eqref{equ2.12a} and \eqref{equ2.12b}. The solvers for matrix
exponential used in FREM integrators are set with default tolerance (machine precision $10^{-16}$). The initial (resp. terminal) conditions for the forward (resp. backward) Lindblad equations are chosen to be
\begin{subequations}\label{equ6.1}
\begin{align}
  &\rho(0)=\frac{1}{2}\left(|0\rangle^{\otimes K}\langle0|^{\otimes K}+
|0\rangle^{\otimes K}\langle d-1|^{\otimes K}+|d-1\rangle^{\otimes K}\langle0|^{\otimes K}+|d-1\rangle^{\otimes K}\langle d-1|^{\otimes K}\right),\label{equ6.1a}\\
&q(T)=\frac{1}{2}\left(|1\rangle^{\otimes K}\langle1|^{\otimes K}+
|1\rangle^{\otimes K}\langle d-2|^{\otimes K}+|d-2\rangle^{\otimes K}\langle1|^{\otimes K}+|d-2\rangle^{\otimes K}\langle d-2|^{\otimes K}\right).\label{equ6.1b}
\end{align}
\end{subequations}
Figures \ref{fig6.1} display the error behavior of the FREM schemes \eqref{equ2.12a} and \eqref{equ2.12b} for the forward and backward Lindblad equations, respectively. We observe the second-order of convergence for the FREM schemes. This is consistent with the convergence results given in Theorems \ref{thm3.2} and \ref{thm4.2}. We can see from Figure \ref{fig6.2} that our FREM schemes preserve positivity and unit trace of the density matrix. Note that we only display evolutions of populations $\rho_{8,8}$ and $q_{4,4}$ for ease of presentation, the positive features can be seen in all populations.

\emph{Performance of the LREM schemes.} We next investigate the convergence behavior of the LREM schemes \eqref{equ2.15} and \eqref{equ2.18}. The initial (resp. terminal) condition of the forward (resp. backward) Lindblad equation is set to be
\begin{align*}
  &\rho(0)=\left(1-\frac{\delta}{2}\right)z_1z_1^\top+\frac{\delta}{2}z_2z_2^\top,\\
  &q(T)=\left(1-\frac{\delta}{2}\right)z_3z_3^\top+\frac{\delta}{2}z_4z_4^\top,
\end{align*}
where $z_1$, $z_2$, $z_3$ and $z_4$ are orthonormal vectors and we obtain them from SVD
of a random $m\times 4$ real matrix. The low-rank initial (resp. terminal) factor is set to $X_0=z_1$ (resp. $Y_N=z_3$) and clearly we have $\|\rho(0)-X_0X_0^{\dag}\|_1=\delta$ (resp. $\|q(T)-Y_NY_N^{\dag}\|_1=\delta$).

\begin{figure}
{ \centering
\includegraphics[width=0.48\textwidth]{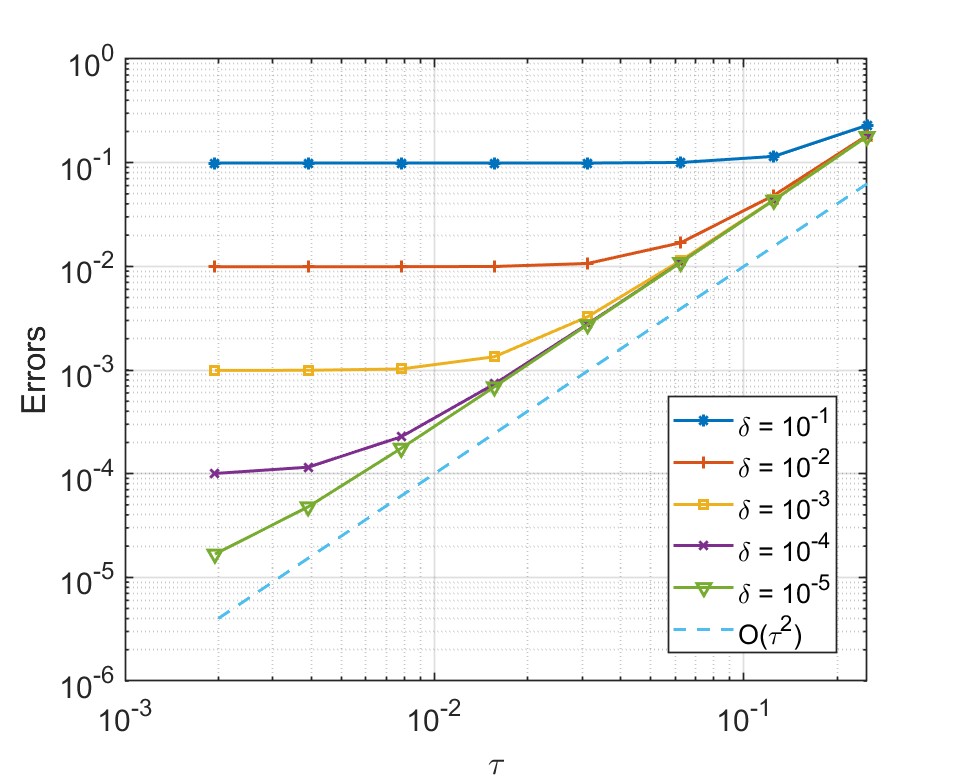}
}  { \centering
\includegraphics[width=0.48\textwidth]{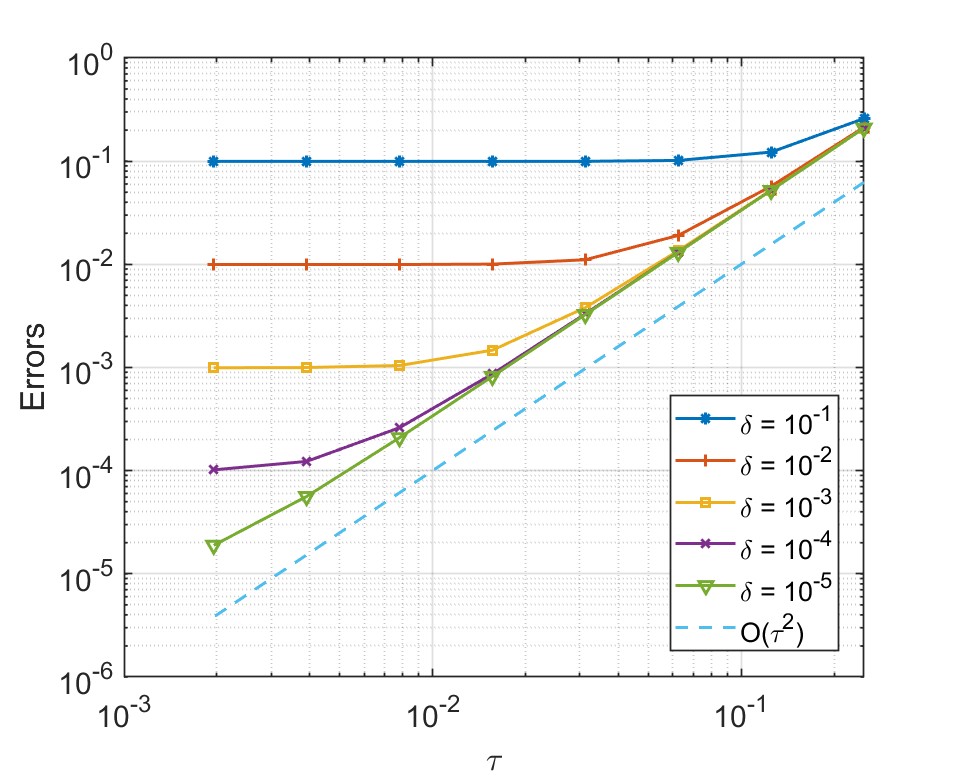}
} \caption{Numerical results of LREM schemes with fixed $\varepsilon_1=\varepsilon_2=10^{-10}$ and different $\delta$ for the Lindblad equations with $d=4$, $K=4$, $a=1.5$, $b=1$, $\gamma=0.05$, $T=1$, $u(t)=\sin(2\pi t)$. Left: errors vs step sizes for the forward Lindblad equation. Right: errors vs step sizes for the backward Lindblad equation.}
\label{fig6.3}
\end{figure}

\begin{figure}
{ \centering
\includegraphics[width=0.48\textwidth]{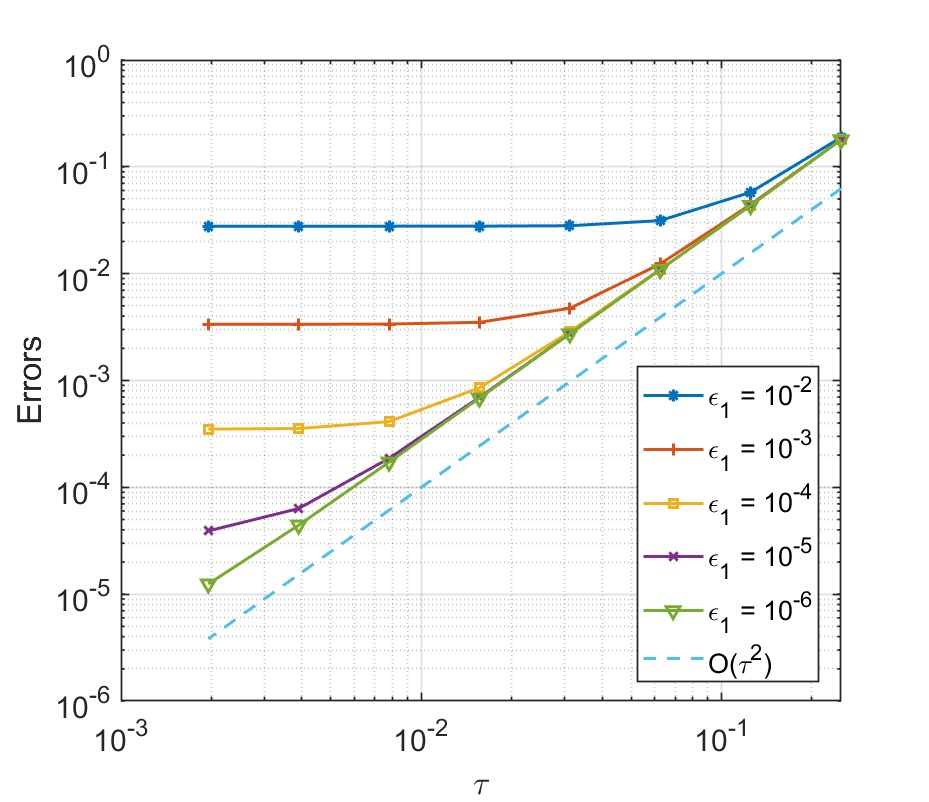}
}  { \centering
\includegraphics[width=0.48\textwidth]{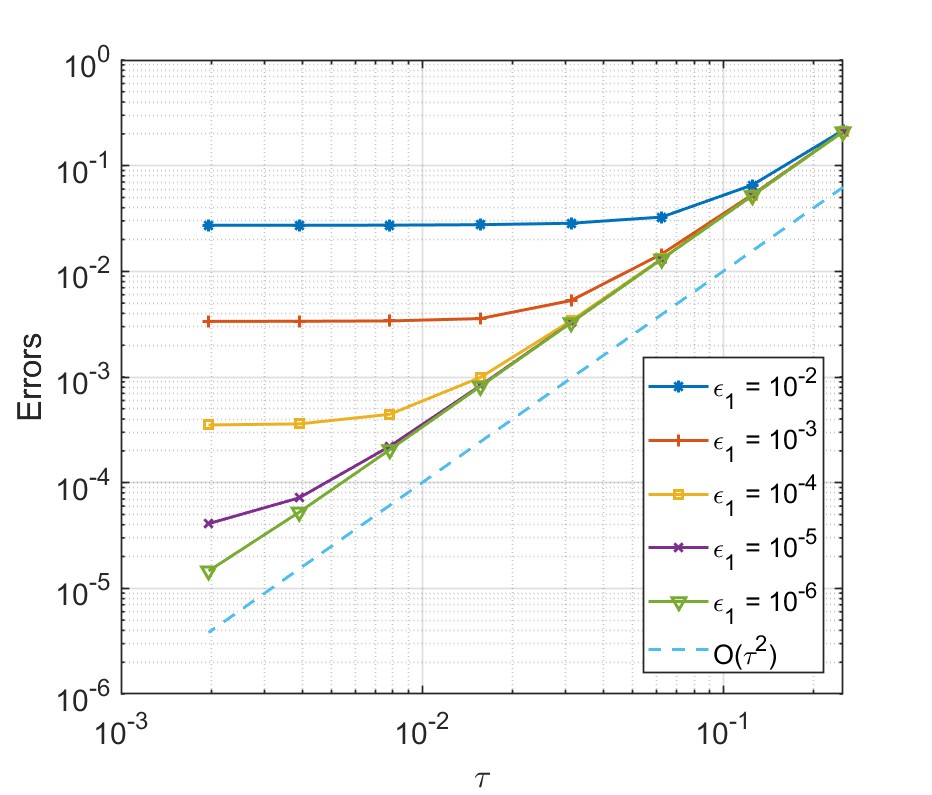}
} \caption{Numerical results of LREM schemes with fixed $\delta=\varepsilon_2=10^{-10}$ and different $\varepsilon_1=\tau\epsilon_1$ for the Lindblad equations with $d=4$, $K=4$, $a=1.5$, $b=1$, $\gamma=0.05$, $T=1$, $u(t)=\sin(2\pi t)$. Left: errors vs step sizes for the forward Lindblad equation. Right: errors vs step sizes for the backward Lindblad equation.}
\label{fig6.4}
\end{figure}

\begin{figure}
{ \centering
\includegraphics[width=0.48\textwidth]{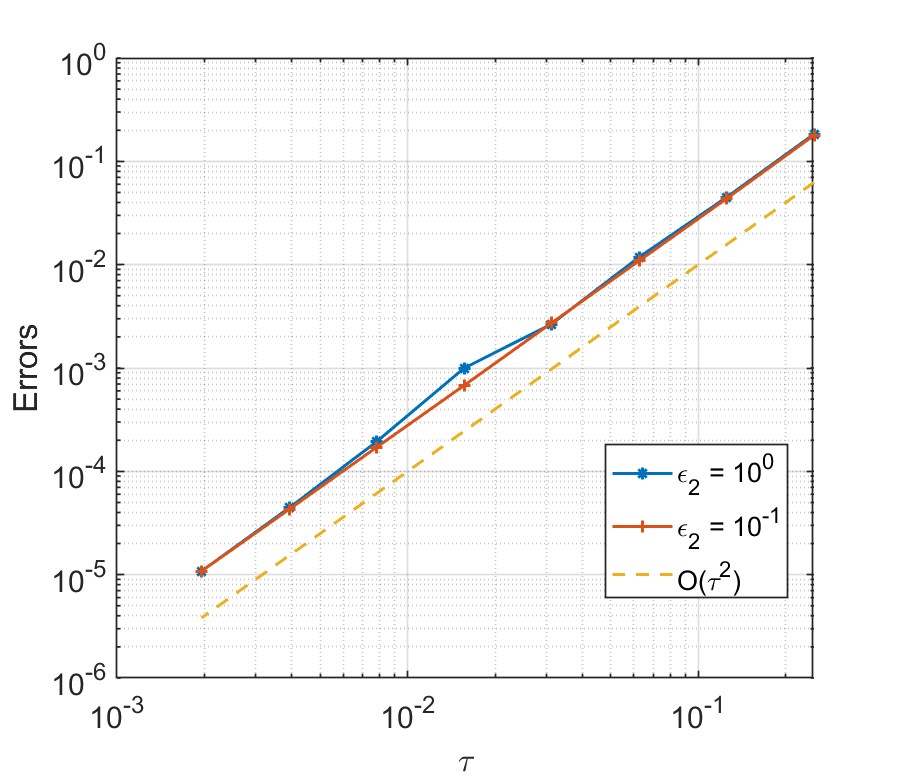}
}  { \centering
\includegraphics[width=0.48\textwidth]{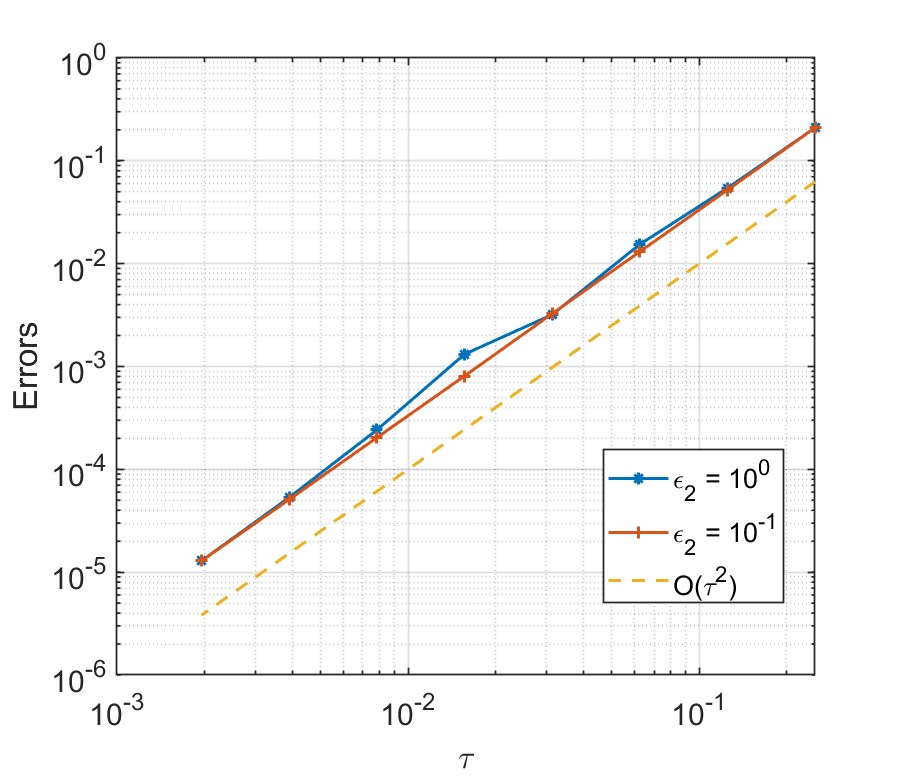}
} \caption{Numerical results of LREM schemes with fixed $\delta=\varepsilon_1=10^{-10}$ and different $\varepsilon_2=\tau\epsilon_2$ for the Lindblad equations with $d=4$, $K=4$, $a=1.5$, $b=1$, $\gamma=0.05$, $T=1$, $u(t)=\sin(2\pi t)$. Left: errors vs step sizes for the forward Lindblad equation. Right: errors vs step sizes for the backward Lindblad equation.}
\label{fig6.5}
\end{figure}

\begin{figure}
{ \centering
\includegraphics[width=0.48\textwidth]{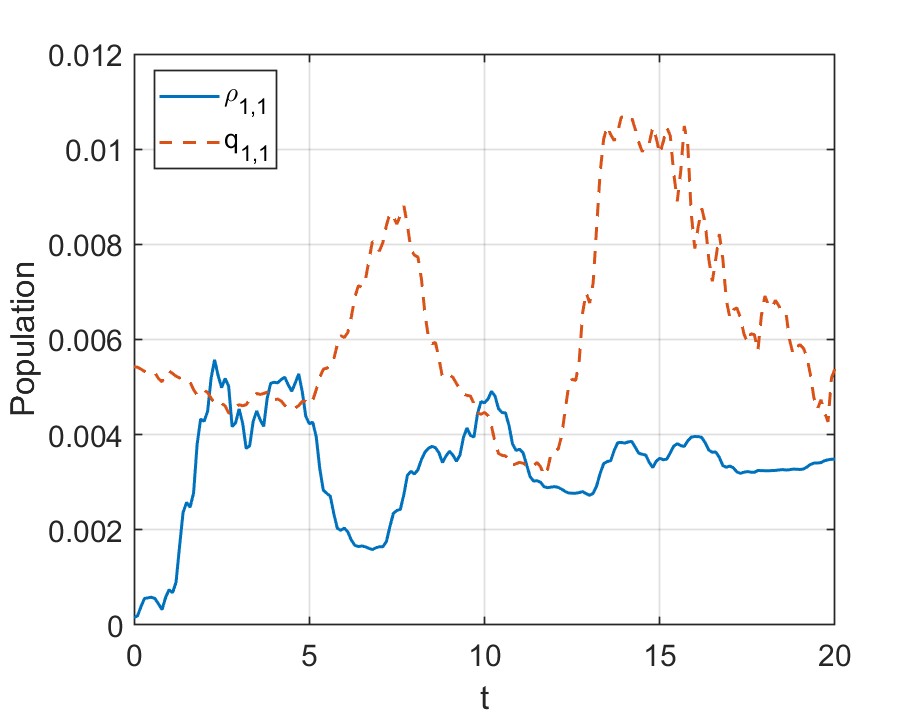}
}  { \centering
\includegraphics[width=0.48\textwidth]{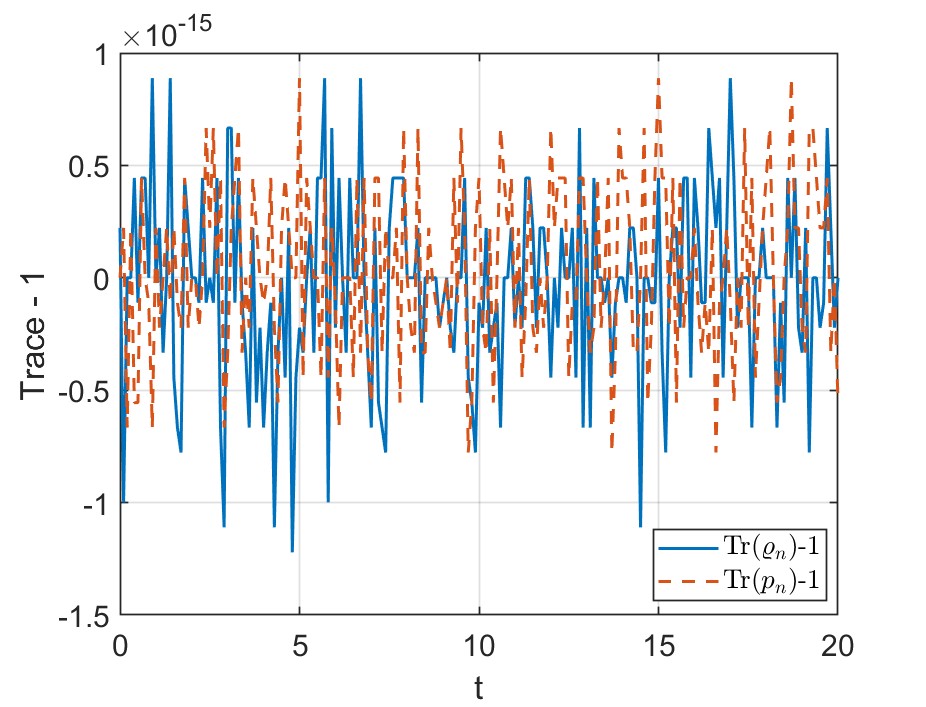}
} \caption{Numerical results of LREM schemes for the Lindblad equations with $d=4$, $K=4$, $a=1.5$, $b=1$, $\gamma=0.05$, $T=20$, $u(t)=\sin(2\pi t)$. Left: evolutions of the populations $\rho_{1,1}$ and $q_{1,1}$ with $\tau=0.1$. Right: evolutions of $\Tr(\varrho_n)-1$ and $\Tr(p_n)-1$.}
\label{fig6.6}
\end{figure}

Figures \ref{fig6.3}-\ref{fig6.5} illustrate the error behavior of the LREM schemes with respect to the initial (resp. terminal) low-rank error, column compression error and matrix exponential error, respectively. Note that the LREM schemes are second-order convergent when the low-rank error and matrix exponential tolerance are small enough. When the low-rank error is dominant, decreasing step size will not lead to smaller global error. These results are consistent with the convergence results given in Theorems \ref{thm3.3} and \ref{thm4.3}. Figures \ref{fig6.6} display evolutions of populations $\rho_{1,1}$, $q_{1,1}$ and traces $\Tr(\varrho_n)-1$, $\Tr(p_n)-1$, from which we observe that our LREM schemes are positivity and trace preserving.

\emph{Comparisons with other method.} We compare the FREM and LREM methods with 
the Lindblad equation solver \emph{mesolve} developed in QuTip \cite{Johansson}. 
We consider the ODE solver \emph{dop853} in \emph{mesolve}, which is based on Dormand and Prince's eighth-order Runge-Kutta method. Note that
the ODE solver \emph{dop853} first reformulates the Lindblad equation in the vectorized form and then integrates. As verified in \cite{Chen3}, QuTip solver \emph{mesolve} does not preserve positive property of the density matrix.

\begin{figure}
{ \centering
\includegraphics[width=0.48\textwidth]{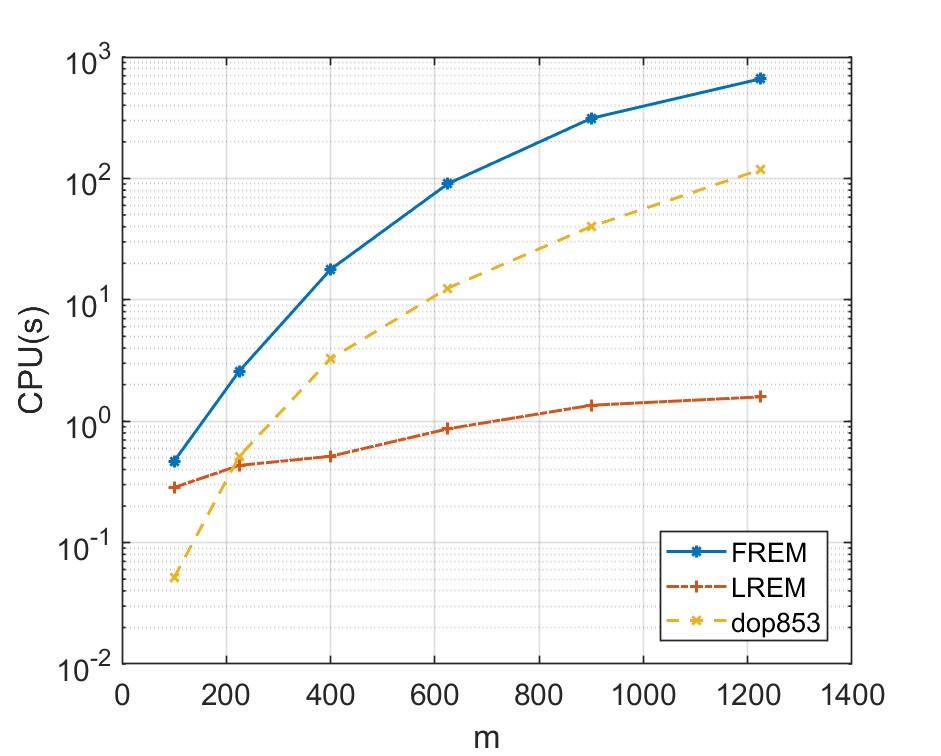}
} { \centering
\includegraphics[width=0.48\textwidth]{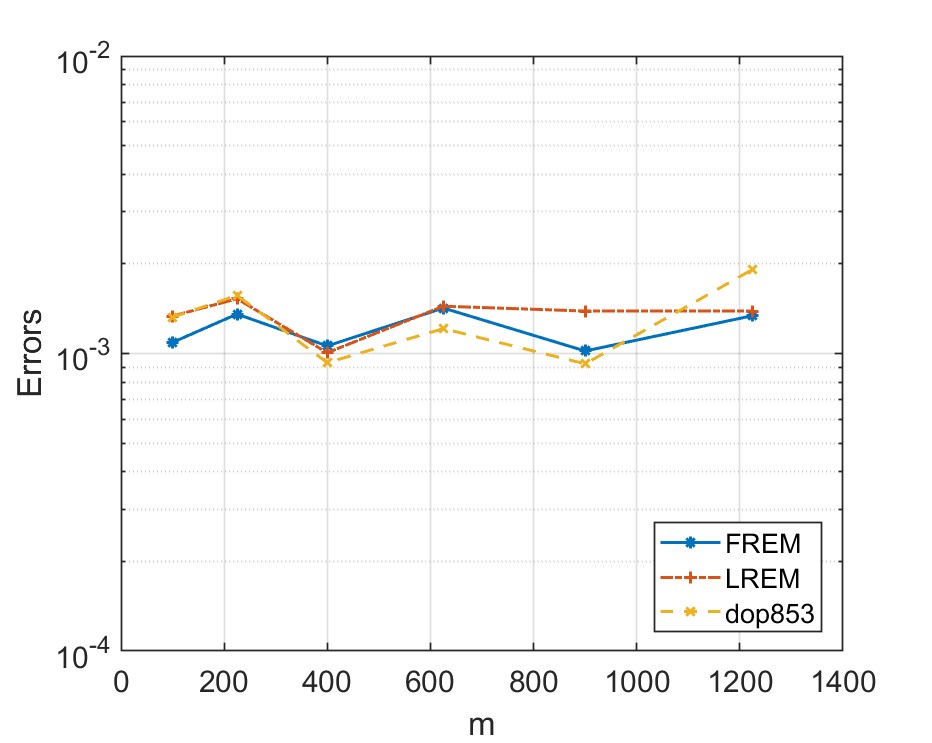}
} \caption{Numerical comparison between the proposed exponential schemes and the QuTip solver for the forward Lindblad equation with $K=2$, $a=1.5$, $b=1$, $\gamma=0.05$, $T=1$, $u(t)=\sin(2\pi t)$. Left: CPU times vs $m$. Right: errors vs $m$.
}
\label{fig6.7}
\end{figure}

\begin{figure}
{ \centering
\includegraphics[width=0.48\textwidth]{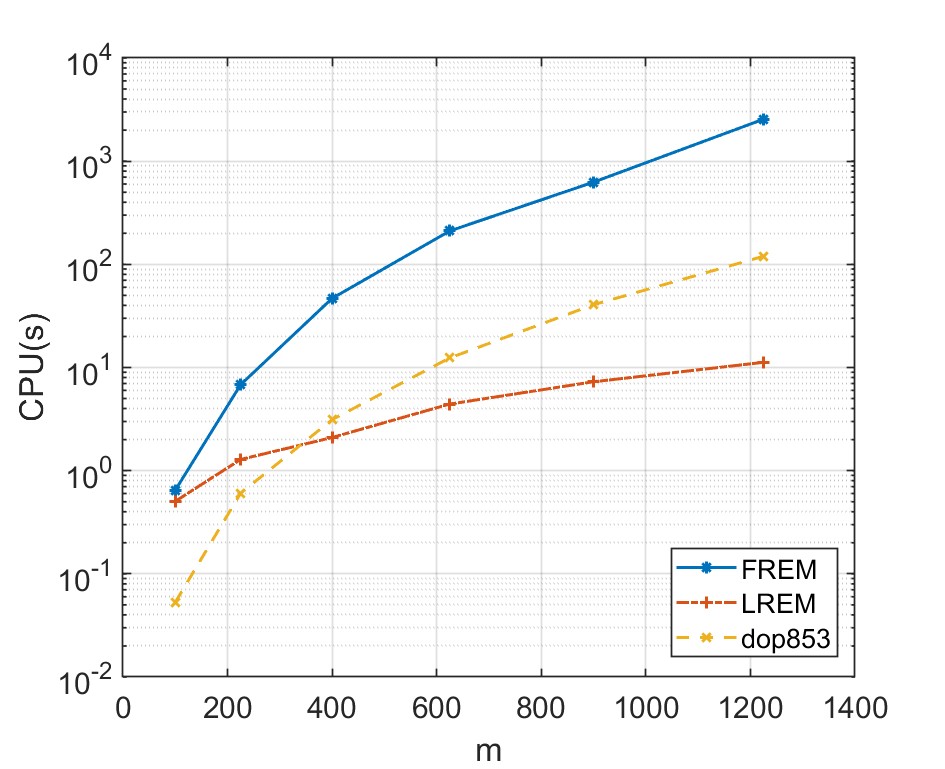}
} { \centering
\includegraphics[width=0.48\textwidth]{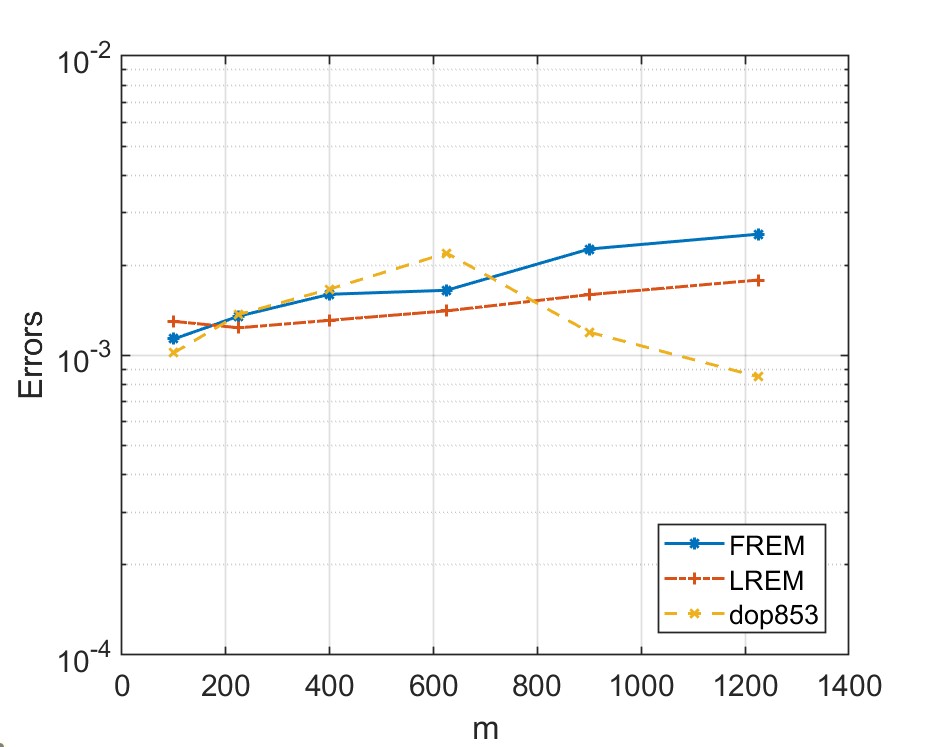}
} \caption{Numerical comparison between the proposed exponential schemes and the QuTip solver for the backward Lindblad equation with $K=2$, $a=1.5$, $b=1$, $\gamma=0.05$, $T=1$, $u(t)=\sin(2\pi t)$. Left: CPU times vs $m$. Right: errors vs $m$.
}
\label{fig6.8}
\end{figure}

In Figures \ref{fig6.7}-\ref{fig6.8}, we compare the computational times (measured in seconds) of our FREM and LREM methods with the solver \emph{dop853} as the size $m=d^K$ of the density matrix increases. The initial and terminal conditions are chosen as in \eqref{equ6.1}. We set $\delta=0$, $\varepsilon_1=\tau^3$ and $\varepsilon_2=10^{-6}$ for the LREM scheme. We choose the absolute and relative tolerances of the QuTip solver and the step sizes of the FREM and LREM schemes such that the errors are approximately $10^{-3}$, see the right-hand plots in Figures \ref{fig6.7}-\ref{fig6.8}. As expected, the LREM scheme performs much faster than the FREM scheme. We also observe that at the same level of accuracy, our LREM scheme is more efficient than the QuTip  Lindblad solver for problems with high dimensions.

\section{Conclusions}\label{sec-conclusions}
We have developed full- and low-rank exponential midpoint integrators for solving the forward and adjoint Lindblad equations. The proposed schemes are shown to preserve positivity and trace unconditionally. Error estimates of these schemes are proved theoretically and verified numerically. We stress that our method could be applied to gradient-based approaches for optimal control of open quantum systems, this will be one of our future works.

\medskip

\def\baselinestretch{0.95}
{\small
\bibliographystyle{abbrv}
\bibliography{literature}

\begin{thebibliography}{10}

\bibitem{Abdelhafez2019}
M.~Abdelhafez, D.~I. Schuster, and J.~Koch.
\newblock Gradient-based optimal control of open quantum systems using quantum
  trajectories and automatic differentiation.
\newblock {\em Phys. Rev. A}, 99:052327, May 2019.

\bibitem{Appelo}
D.~Appelö and Y.~Cheng.
\newblock Kraus is king: High-order completely positive and trace preserving
  ({CPTP}) low rank method for the {L}indblad master equation.
\newblock {\em Journal of Computational Physics}, 534:114036, 2025.

\bibitem{Bidegaray}
B.~Bidégaray, A.~Bourgeade, and D.~Reignier.
\newblock Introducing physical relaxation terms in {B}loch equations.
\newblock {\em Journal of Computational Physics}, 170(2):603--613, 2001.

\bibitem{Borzi2017}
A.~Borz{\`\i}, G.~Ciaramella, and M.~Sprengel.
\newblock {\em Formulation and Numerical Solution of Quantum Control Problems}.
\newblock Society for Industrial and Applied Mathematics, Philadelphia, PA,
  2017.

\bibitem{Saut}
A.~Bourgeade and O.~Saut.
\newblock Numerical methods for the bidimensional {M}axwell-{B}loch equations
  in nonlinear crystals.
\newblock {\em Journal of Computational Physics}, 213(2):823--843, 2006.

\bibitem{Boutin2017}
S.~Boutin, C.~K. Andersen, J.~Venkatraman, A.~J. Ferris, and A.~Blais.
\newblock Resonator reset in circuit qed by optimal control for large open
  quantum systems.
\newblock {\em Phys. Rev. A}, 96:042315, Oct 2017.

\bibitem{Breuer}
H.-P. Breuer and F.~Petruccione.
\newblock {\em {The Theory of Open Quantum Systems}}.
\newblock Oxford University Press, 2007.

\bibitem{Caneva2011}
T.~Caneva, T.~Calarco, and S.~Montangero.
\newblock Chopped random-basis quantum optimization.
\newblock {\em Phys. Rev. A}, 84:022326, Aug 2011.

\bibitem{Cao}
Y.~Cao and J.~Lu.
\newblock Structure-preserving numerical schemes for {L}indblad equations.
\newblock {\em Journal of Scientific Computing}, 102:27, 2025.

\bibitem{Chen3}
H.~Chen, A.~Borz{\`\i}, D.~Jankovi\'{c}, J.-G. Hartmann, and P.-A. Hervieux.
\newblock Full- and low-rank exponential euler integrators for the lindblad
  equation, 2024.

\bibitem{Alessandro2008}
D.~D'~Alessandro.
\newblock {\em Introduction to Quantum Control and Dynamics}.
\newblock Chapman and Hall/CRC, 2008.

\bibitem{Davies1976}
E.~Davies.
\newblock {\em Quantum Theory of Open Systems}.
\newblock Academic Press, 1976.

\bibitem{Fouquieres2011}
P.~{de Fouquieres}, S.~Schirmer, S.~Glaser, and I.~Kuprov.
\newblock Second order gradient ascent pulse engineering.
\newblock {\em Journal of Magnetic Resonance}, 212(2):412--417, 2011.

\bibitem{Doria2011}
P.~Doria, T.~Calarco, and S.~Montangero.
\newblock Optimal control technique for many-body quantum dynamics.
\newblock {\em Phys. Rev. Lett.}, 106:190501, May 2011.

\bibitem{Egger2014}
D.~J. Egger and F.~K. Wilhelm.
\newblock Optimal control of a quantum measurement.
\newblock {\em Phys. Rev. A}, 90:052331, Nov 2014.

\bibitem{Goerz2014}
M.~Goerz, D.~Reich, and C.~Koch.
\newblock Optimal control theory for a unitary operation under dissipative
  evolution.
\newblock {\em New Journal of Physics}, 16(5):055012, may 2014.

\bibitem{Gollub2008}
C.~Gollub, M.~Kowalewski, and R.~de~Vivie-Riedle.
\newblock Monotonic convergent optimal control theory with strict limitations
  on the spectrum of optimized laser fields.
\newblock {\em Phys. Rev. Lett.}, 101:073002, Aug 2008.

\bibitem{Gorini}
V.~Gorini, A.~Kossakowski, and E.~Sudarshan.
\newblock {Completely positive dynamical semigroups of N-level systems}.
\newblock {\em Journal of Mathematical Physics}, 17(5):821--825, 05 1976.

\bibitem{Johansson}
J.~Johansson, P.~Nation, and F.~Nori.
\newblock Qutip 2: A {P}ython framework for the dynamics of open quantum
  systems.
\newblock {\em Computer Physics Communications}, 184(4):1234--1240, 2013.

\bibitem{Khaneja2001}
N.~Khaneja, R.~Brockett, and S.~J. Glaser.
\newblock Time optimal control in spin systems.
\newblock {\em Phys. Rev. A}, 63:032308, Feb 2001.

\bibitem{Khaneja2005}
N.~Khaneja, T.~Reiss, C.~Kehlet, T.~Schulte-Herbrüggen, and S.~J.~Glaser.
\newblock Optimal control of coupled spin dynamics: design of nmr pulse
  sequences by gradient ascent algorithms.
\newblock {\em Journal of Magnetic Resonance}, 172(2):296--305, 2005.

\bibitem{Krotov1995}
V.~Krotov.
\newblock {\em Global Methods in Optimal Control Theory}.
\newblock CRC Press, 1995.

\bibitem{LeBris1}
C.~Le~Bris and P.~Rouchon.
\newblock Low-rank numerical approximations for high-dimensional {L}indblad
  equations.
\newblock {\em Phys. Rev. A}, 87:022125, Feb 2013.

\bibitem{LeBris2}
C.~Le~Bris, P.~Rouchon, and J.~Roussel.
\newblock Adaptive low-rank approximation and denoised {M}onte {C}arlo approach
  for high-dimensional {L}indblad equations.
\newblock {\em Phys. Rev. A}, 92:062126, Dec 2015.

\bibitem{Lindblad}
G.~Lindblad.
\newblock On the generators of quantum dynamical semigroups.
\newblock {\em Communications in Mathematical Physics}, 48(2):119--130, 1976.

\bibitem{Machnes2011}
S.~Machnes, U.~Sander, S.~J. Glaser, P.~de~Fouqui\`eres, A.~Gruslys,
  S.~Schirmer, and T.~Schulte-Herbr\"uggen.
\newblock Comparing, optimizing, and benchmarking quantum-control algorithms in
  a unifying programming framework.
\newblock {\em Phys. Rev. A}, 84:022305, Aug 2011.

\bibitem{Maday2003}
Y.~Maday and G.~Turinici.
\newblock New formulations of monotonically convergent quantum control
  algorithms.
\newblock {\em The Journal of Chemical Physics}, 118(18):8191--8196, 05 2003.

\bibitem{Ohtsuki2004}
Y.~Ohtsuki, G.~Turinici, and H.~Rabitz.
\newblock Generalized monotonically convergent algorithms for solving quantum
  optimal control problems.
\newblock {\em The Journal of Chemical Physics}, 120(12):5509--5517, 03 2004.

\bibitem{Palao2002}
J.~Palao and R.~Kosloff.
\newblock Quantum computing by an optimal control algorithm for unitary
  transformations.
\newblock {\em Phys. Rev. Lett.}, 89:188301, Oct 2002.

\bibitem{Pechen2006}
A.~Pechen and H.~Rabitz.
\newblock Teaching the environment to control quantum systems.
\newblock {\em Phys. Rev. A}, 73:062102, Jun 2006.

\bibitem{Rach2015}
N.~Rach, M.~M. M\"uller, T.~Calarco, and S.~Montangero.
\newblock Dressing the chopped-random-basis optimization: A bandwidth-limited
  access to the trap-free landscape.
\newblock {\em Phys. Rev. A}, 92:062343, Dec 2015.

\bibitem{Reich2012}
D.~M. Reich, M.~Ndong, and C.~P. Koch.
\newblock Monotonically convergent optimization in quantum control using
  krotov's method.
\newblock {\em The Journal of Chemical Physics}, 136(10):104103, 03 2012.

\bibitem{Riesch}
M.~Riesch and C.~Jirauschek.
\newblock Analyzing the positivity preservation of numerical methods for the
  {L}iouville-von {N}eumann equation.
\newblock {\em Journal of Computational Physics}, 390:290--296, 2019.

\bibitem{Riesch1}
M.~Riesch, A.~Pikl, and C.~Jirauschek.
\newblock Completely positive trace preserving methods for the {L}indblad
  equation.
\newblock In {\em 2020 International Conference on Numerical Simulation of
  Optoelectronic Devices (NUSOD)}, pages 109--110, 2020.

\bibitem{Schlimgen}
A.~Schlimgen, K.~Head-Marsden, L.~Sager, P.~Narang, and D.~Mazziotti.
\newblock Quantum simulation of the {L}indblad equation using a unitary
  decomposition of operators.
\newblock {\em Phys. Rev. Res.}, 4:023216, Jun 2022.

\bibitem{Herbruggen2005}
T.~Schulte-Herbr\"uggen, A.~Sp\"orl, N.~Khaneja, and S.~J. Glaser.
\newblock Optimal control-based efficient synthesis of building blocks of
  quantum algorithms: A perspective from network complexity towards time
  complexity.
\newblock {\em Phys. Rev. A}, 72:042331, Oct 2005.

\bibitem{Schulte-Herbruggen2011}
T.~Schulte-Herbrüggen, A.~Sporl, N.~Khaneja, and S.~Glaser.
\newblock Optimal control for generating quantum gates in open dissipative
  systems.
\newblock {\em Journal of Physics B: Atomic, Molecular and Optical Physics},
  44(15):154013, jul 2011.

\bibitem{TOSNER2009}
Z.~T\v{o}sner, T.~Vosegaard, C.~Kehlet, N.~Khaneja, S.~J.~Glaser, and
  N.~Nielsen.
\newblock Optimal control in nmr spectroscopy: Numerical implementation in
  simpson.
\newblock {\em Journal of Magnetic Resonance}, 197(2):120--134, 2009.

\bibitem{Wenin2008}
M.~Wenin and W.~P\"otz.
\newblock State-independent control theory for weakly dissipative quantum
  systems.
\newblock {\em Phys. Rev. A}, 78:012358, Jul 2008.

\bibitem{Wiseman2009}
H.~Wiseman and G.~Milburn.
\newblock {\em Quantum Measurement and Control}.
\newblock Cambridge University Press, 2009.

\bibitem{Zhu1998}
W.~Zhu, J.~Botina, and H.~Rabitz.
\newblock Rapidly convergent iteration methods for quantum optimal control of
  population.
\newblock {\em The Journal of Chemical Physics}, 108(5):1953--1963, 02 1998.

\bibitem{Ziolkowski}
R.~Ziolkowski, J.~Arnold, and D.~Gogny.
\newblock Ultrafast pulse interactions with two-level atoms.
\newblock {\em Phys. Rev. A}, 52:3082--3094, Oct 1995.

\end{thebibliography}

}
\end{document}